\documentclass[final]{article}

\usepackage{arxiv}

\usepackage[utf8]{inputenc} 
\usepackage[T1]{fontenc}    
\usepackage{hyperref}       
\usepackage{url}            
\usepackage{booktabs}       
\usepackage{amsfonts}       
\usepackage{nicefrac}       
\usepackage{microtype}      
\usepackage{lipsum}
\usepackage{graphicx}

\usepackage{xcolor}
\usepackage{tikz}
\usepackage{amsmath}
\usepackage{amssymb}

\usepackage{lineno}
\usepackage{algorithm2e}
\usepackage{textcomp} 
\usepackage{subcaption}
\usepackage{adjustbox}

\modulolinenumbers[5]
\usetikzlibrary{calc}
\usetikzlibrary{intersections}
\usetikzlibrary{through}

\title{Multi-Layer Wind Velocity Field Visualization in Infrared Images of Clouds for Solar Irradiance Forecasting}

\author{
 Guillermo Terr\'en-Serrano \\
  Department of Electrical and Computer Engineering \\
  The University of New Mexico \\
  Albuquerque, NM 87131, United States\\
  \texttt{guillermoterren@unm.edu} \\
 \And
  Manel Mart\'inez-Ram\'on \\
  Department of Electrical and Computer Engineering \\
  The University of New Mexico \\
  Albuquerque, NM 87131, United States\\
  \texttt{manel@unm.edu} \\
}

\begin{document}

\maketitle

\begin{abstract}
    The energy available in a solar energy powered grid is uncertain due to the weather conditions at the time of generation. Forecasting global solar irradiance could address this problem by providing the power grid with the capability of scheduling the storage and dispatch of energy. The occlusion of the Sun by clouds is the main cause of instabilities in the generation of solar energy. This investigation proposes a method to visualize the wind velocity field in sequences of longwave infrared images of clouds when there are multiple wind velocity fields in an image. This method can be used to forecast the occlusion of the Sun by clouds, providing stability in the generation of solar energy. Unsupervised learning is implemented to infer the distribution of the clouds' velocity vectors and heights in multiple wind velocity fields in an infrared image. A multi-output weighted support vector machine with flow constraints is used to extrapolate the wind velocity fields to the entire frame, visualizing the path of the clouds. The proposed method is capable of approximating the wind velocity field in a small air parcel using the velocity vectors and physical features of clouds extracted from infrared images. Assuming that the streamlines are pathlines, the visualization of the wind velocity field can be used for forecasting cloud occlusions of the Sun. This is of importance when considering ways of increasing the stability of solar energy generation.
\end{abstract}


\section{Introduction}

Recent legislative initiatives to stimulate the use of solar power and other sustainable energy sources will increase the number of solar power plants connected to urban power grids worldwide \cite{HALU2016}. California aims to have 100\%  of clean energy generation by 2045 \cite{CA}. Similar initiatives are occurring in Japan, South Africa and the European Union, where local governments aim to generate the 24\% \cite{JP}, 41\% \cite{SA}, and 32\% \cite{EU} of their energy from renewable sources by 2030 respectively. In addition, the growth of Photovoltaic (PV) solar power capacity has continued to increase in a steady exponential scale from 2000 \cite{PVgrowth}.
 
To increase the percentage of solar energy in the electrical power grid it is important to guarantee a reliable supply of energy \cite{BEYEA2010}. The forecasting of solar power provides a Smart Grid (SG) with the capability of performing energy management \cite{WAN2015}. The interruptions in energy supply from PV systems occurs due to the projection of shadows from passing clouds \cite{CHEN2020}. Moving clouds have different effects depending on the configuration of PV arrays \cite{LAPPALAINEN2017}, and may cause the solar irradiance received by a PV system to increase or decrease \cite{MATEOS2014}. The forecasting of solar irradiance in microgrids \cite{DIAGNE2013} allows automatic control of home appliances and other devices \cite{CHOU2019}. In a large-scale SG with a stable supply of energy using a mix plan of solar power from PV, concentrated solar power \cite{CRESPI2018} and other systems, the forecasting of solar energy is necessary to perform an efficient management of the resources \cite{YANG2018}.

There is a documented relationship between ground measurements of direct normal irradiance and Cloud Index (CI) \cite{FURLAN2012}. The relationship holds in diverse climates and weather conditions \cite{ESCRIG2013, PEREZ2010} when the CI is calculated from visible and infrared (IR) light sensors mounted in geostationary satellites \cite{ARBIZU2017, HAMMER1999, INEICHEN1999, SCHILLINGS2004}. On-ground maps of solar irradiance can be derived from the CI using geostationary satellite images \cite{JIANG2020, PRASAD2015}. 

This research aims to visualize the wind velocity field to anticipate interruptions in the supply of energy generated by PV systems \cite{PEDRO2012}. The forecasting interval of this application is from 1 to 5 minutes \cite{MAMMOLI2013}. This is often called nowcasting \cite{VOYANT2017}. Accurate Numerical Weather Prediction (NWP) models which analyze atmospheric dynamics using satellite images are computationally expensive due to the resolution of the numerical grid necessary to analyze the forecasting intervals \cite{KALNAY2003}. The variables in mesoscale meteorology models have collinearity when the objective is to forecast solar irradiation \cite{GARCIA2018}.

When transmitting images from a satellite, communications might have a delay of up to an hour \cite{JANG2016}. An alternative to satellite cameras is the total sky imager \cite{GOHARI2014}. This device captures sky images reflected on a concave mirror, and allows a high Field Of View (FOV) \cite{TSI2001}. This device has a number of disadvantages including its cost, and the projection of shadows on the mirror from objects in its own structure \cite{MARQUEZ2013}. Digital cameras are less expensive and can obtain better performances around the circumsolar area with attached lenses \cite{KUHN2017}. In fact, near IR filters can attenuate the scattering produced by solar irradiance \cite{MAMMOLI2013}, and fish-eye lenses increase the FOV, which allows recording of low-cost, shadow-free sky images \cite{CHENG2017}. IR sensors are the most viable alternative when the forecast is meant for hours ahead, and includes night hours or poor daylight conditions \cite{DENEKE2008}. Recent technological innovations have increased the FOV of ground-based IR images \cite{MAMMOLI2019}.

The visualization of the wind velocity field requires measurements of wind velocity at a given altitude. The wind velocity increases with the altitude in the lower atmosphere \cite{WIZELIUS2007}. The decrease of temperature along the Troposphere can be approximated by a linear function \cite{GLICKMAN2000}. Cloud formations are feasible in a range of altitudes that varies from the ground to the Tropopause \cite{RANDEL2000}. The detection of clouds in IR images allows us to indirectly measure physical magnitudes of the wind velocity field \cite{SHAW2005}. Radiometric IR cameras provide uniform thermal imaging \cite{SHAW2013}, and may be stabilized to perform atmospheric measurements \cite{NUGENT2013}. In fact, microbolometer IR cameras have been used to provide statistical analysis of clouds \cite{THURAIRAJAH2007} for Earth-space communication \cite{NUGENT2009}.

Methods of computational numerical analysis are an effective way to analyze images of clouds. The direction and magnitude of cloud velocity have been estimated applying motion vector techniques to a series of consecutive frames \cite{CHOW2011}. Through image segmentation, it is possible to identify clouds and other objects in an image \cite{KUHN2017}. The clouds' pathlines can be estimated by tracking them with a Kalman filter \cite{CHENG2017}. Classical methods of statistical modeling and linear regression have low computational requirements, and are an alternative to complex NWP models \cite{FU2013}. Machine learning (ML) algorithms such as artificial neural networks \cite{KONG2020}, or Support Vector Machines (SVM) \cite{DENG2019}, are promising models to find space-time correlations in cloud images. 

This research utilizes two innovations. First, a Data Acquisition (DAQ) system is used for capturing radiometric long-wave IR circumsolar images combined with pyranometer measurements \cite{TERREN2020a}. The DAQ is equipped with a solar tracker that updates its pan and tilt every second, maintaining the Sun in a central position in the images during the day \cite{REDA2004}. The IR images are taken at an angle from the normal position of the camera in relation to the ground. The angle is the Sun's elevation. This causes the relative distance of a given object on the horizon to increase from top to bottom in an image. To account for this effect, a second innovation is introduced to transform the velocity vectors from the original Euclidean frame of reference to a non-linear frame of reference \cite{TERREN2020a}.

This research also proposes and implements an online ML algorithm for predicting the streamlines of multiple wind flows in an image. An unsupervised ML algorithm infers the distribution of velocity vectors and heights of multiple layers of clouds. The velocity vectors are approximated using the Weighted Lucas-Kanade (WLK) method, and are segmented and subsampled to reduce the noise of the approximation and the computational burden of the entire algorithm. A Multi-Output Weighted Support Vector Machine ($\varepsilon$-MO-WSVM) \cite{VAPNIK1998} visualizes the approximated velocity vectors to predict the trajectories of the clouds. The $\varepsilon$-MO-WSVM is modified from its original form adding flow constraints. The flow constraints are added so that the approximated streamlines are equivalent to the pathlines. The wind velocity field visualization can be used to forecast occlusion of the Sun by clouds, thereby predicting and preventing disruptions in the generation of energy from solar power plants.

\section{Wind Velocity Field}

The IR sensor produces a uniform thermal image. When the radiometry functionality is enabled, the pixels in a frame are turned into temperature measurements. A pixel of the camera frame is defined by a pair of euclidean coordinates $\mathbf{X} = \{ (x, y )_{i,j} \mid \forall i = 1, \ldots, M, \ \forall j = 1, \ldots, N \}$, and the temperature of each one of the pixels is defined in Kelvin degrees as $ \mathbf{T}^k = \{ T_{i,j} \in \mathbb{R} \mid \forall i = 1, \ldots, M, \ \forall j = 1, \ldots, N \}$, where $k$ represents a process defined as $k \in \left( 0, \infty \right]$, which is a sequence of IR images ordered chronologically by time of acquisition. The temperature of a particle in the Troposphere is a function of the height \cite{MURALIKRISHNA2017}. The height of a pixel in a frame is approximated using the Moist Adiabatic Lapse Rate (MALR) function \cite{TERREN2020a}, that we define as $\phi : T \mapsto h $, knowing the temperatures obtained with the IR camera. The height of each one of the pixels in a frame are $\mathbf{H}^k = \{ H_{i,j} \in \mathbb{R} \mid \forall i = 1, \ldots, M, \ \forall j = 1, \ldots, N \}$.

When there are multiple layers of clouds in an image, a Beta Mixture Model (BeMM) of the temperature of the pixels is expected to have multiple clusters. The number of clusters $C$ is estimated by a previously trained detection algorithm that infers the number of wind velocity fields which are in an image. In order to infer the distribution of the temperature of the pixels with a BeMM, the temperatures are first normalized to the domain of a beta distribution such as $\bar{T}_{i,j} = [ T_{i,j} - \min ( \mathbf{T}^k ) ] / [\max ( \mathbf{T}^k ) - \min ( \mathbf{T}^k ) ]$. 

\subsection{Beta Mixture Model}

Consider the temperatures $\bar{T}_{i,j}$ of a given image (by omitting superindex $k$). The distribution of the normalized temperatures can be approximated by a mixture of beta distributions $\bar{T} \sim Be ( \alpha_c, \beta_c )$ with the density function,
\begin{align}
    \label{eq:beta_distribution}
    f \left( \bar{T}_{i,j} ; \alpha_c, \beta_c \right) = \frac{1}{\mathrm{B} \left( \alpha_c, \beta_c \right)} \cdot \bar{T}_{i,j}^{\alpha_c - 1} \cdot \left( 1 - \bar{T}_{i,j}\right)^{\beta_c - 1}, \quad \alpha_c, \beta_c > 0,
\end{align}
where $\bar{T}_{i,j} \in (0, 1)$,  the beta function is $\mathrm{B} ( \alpha_c, \beta_c ) = [ \Gamma ( \alpha_c ) \Gamma ( \beta_c ) ] / [\Gamma ( \alpha_c + \beta_c ) ] $, and the gamma function is $\Gamma ( \alpha_c ) = ( \alpha_c - 1 ) !$.

The log-likelihood of the beta density function that we need to compute the expected complete data log-likelihood (CDLL) is,
\begin{align}
    \log p \left( \bar{T}_{i,j}\mid \alpha_c, \beta_c \right) = \left( \alpha_c - 1 \right) \log \bar{T}_{i,j}+ \left( \beta_c - 1 \right) \log \left( 1 - \bar{T}_{i,j} \right)- \log \mathrm{B} \left(\alpha_c, \beta_c \right),
\end{align}
and the CDLL in a mixture model is,
\begin{align}\label{EM_likelihood}
   \mathcal{Q} \left( \boldsymbol{\theta}^{\left( t\right)}, \boldsymbol{\theta}^{\left(t - 1\right)}\right) = \sum_{i = 1}^M \sum_{j = 1}^N \sum_{c = 1}^C \gamma_{i,j,c} \log \pi_c + \sum_{i = 1}^M \sum_{j = 1}^N \sum_{c = 1}^C \gamma_{i,j,c}  \log p \left( \bar{T}_{i,j} \mid \boldsymbol{\theta}^{\left( t\right)} \right)
\end{align}
where $\gamma_{i,j,c} \triangleq  p ( y_{i,j} = c \mid \bar{T}_{i,j}, \boldsymbol{\theta}^{(t - 1)} )$ is the responsibility of the cluster $c$ in the sample $i,j$ and $\boldsymbol{\theta}^{( t)} = \{\alpha_c^{\left( t\right)} , \beta_c^{( t)} \}$.

The parameters in the clustering of beta distributions can be directly computed applying the Expectation Maximization (EM) algorithm \cite{MURPHY2012}. In the E stage of the algorithm a prior is established and then, by using the likelihood function \eqref{EM_likelihood}, a posterior $\gamma_{i,j,c} = p(y_{i,j} = c \mid \bar{T}_{i,j}, \boldsymbol{\theta})$ can be assigned to each sample \cite{BISHOP2006}. In the M stage, the parameters $\alpha_c$ and $\beta_c$ of each cluster that maximize the log-likelihood are computed by gradient descent of the CDLL \cite{NOCEDAL2006}. The corresponding derivatives are,
\begin{align}
    \frac{\partial \mathcal{L} \left( \boldsymbol{\theta} \right)}{\partial \alpha_c} &=  \sum_{i = 1}^M \sum_{j = 1}^N \sum_{c = 1}^C \gamma_{i,j,c} \frac{\partial}{\partial \alpha_c} \log p \left( \bar{T}_{i,j} \mid \alpha_c, \beta_c \right) \\
    &= \sum_{i = 1}^M \sum_{j = 1}^N \gamma_{i,j,c} \sum_{c = 1}^C \left[ \log \bar{T}_{i,j} - \psi \left( \alpha_c \right) + \psi \left( \alpha_c + \beta_c \right) \right], \\
    \frac{\partial \mathcal{L} \left( \boldsymbol{\theta}_c \right)}{\partial \beta_c} &= \sum_{i = 1}^M \sum_{j = 1}^N \sum_{c = 1}^C \gamma_{i,j,c} \left[
    \frac{\partial}{\partial \beta_c} \log p \left( \bar{T}_{i,j} \mid \alpha_c, \beta_c \right) \right] \\
    &= \sum_{i = 1}^M \sum_{j = 1}^N \gamma_{i,j,c} \sum_{c = 1}^C \left[ \log \left( - \bar{T}_{i,j} \right)- \psi \left( \beta_k \right) + \psi \left( \alpha_c + \beta_c \right) \right].
\end{align} 
where $\partial \mathrm{B} ( \alpha_c, \beta_c ) / \partial \alpha_c  = \mathrm{B} ( \alpha_c, \beta_c ) \cdot [ \psi ( \alpha_c ) - \psi ( \alpha_c + \beta_c ) ]$, and $\psi ( \cdot )$ is the digamma function, which is $\psi ( \alpha_c ) = \Gamma^\prime ( \alpha_c ) / \Gamma \left( \alpha_c \right)$.

The optimal priors are found by maximizing the CDLL with respect to $\pi_c$, constrained to $\sum_c\pi_c=1$. As a result, the optimal priors are
\begin{equation}
    \pi_c = \frac{1}{MN}\sum_{i = 1}^M \sum_{j = 1}^N \gamma_{i,j,c}.
\end{equation}

The cloud average heights in a frame are computed using the posterior probabilities $\gamma_{i,j,c}$ in a frame, but only in the pixels with a cloud,
\begin{align}
    \hat{H}_c = \frac{\sum_{i,j} \gamma_{i,j,c} \cdot H_{i,j} \cdot \mathbb{I} \left( b_{i,j} = 1 \right)}{\sum_{i,j} \gamma_{i,j,c} \cdot \mathbb{I} \left( b_{i,j} = 1 \right)},
\end{align}
where $\mathbb{I} \left( \cdot \right)$ is the indicator function. An image segmentation algorithm indicates which pixels belong to a cloud, so that $\mathbf{B} = \{ b_{i,j} \in \mathbb{B} \mid \forall i = 1, \ldots, M, \ \forall j = 1, \ldots, N \}$ is a binary image where 0 is a clear sky pixel, and 1 is a pixel belonging to a cloud \cite{TERREN2020c}. 

The performance of a Gamma Mixture Model and a BeMM were compared to infer the distribution of the temperatures and the heights. The BeMM of the temperature of the pixels was found to be better when identifying which pixels belong to the wind velocity layer.

\subsection{Motion Vectors}

In current computer vision literature, there are three primary methods to estimate the motion of objects in a sequence of images: Lucas-Kanade \cite{LUCAS1981}, Horn-Schunk \cite{HORN1981}, and Farneb{\"a}ck \cite{FERNEBACK2003}. These three methods are based on the space-time partial derivatives between two consecutive frames. Taking a different disciplinary approach, the velocity field in experimental fluid dynamics is approximated applying research methods based on signal cross-correlation operated in the frequency domain \cite{ADRIAN2011}. The techniques to estimate the motion vectors in an image are sensitive to the intensity gradient of the pixels. We implemented a model that removes the gradient produced by the solar direct radiation, and atmospheric scattered radiation, both of which routinely appear on the images in the course of the year. A persistent model of the outdoor germanium window of the camera removes sporadic debris that appears in the images such as water stains or dust particles \cite{TERREN2020a}.

A series of sequences of images with clouds flowing in different directions were simulated to cross-validate the set of parameters for each one of the mentioned methods. The investigation searched for a dense implementation of a motion vector method to approximate the dynamics of clouds. The most suitable method was found to be the Weighted Lucas-Kanande (WLK) \cite{SIMON2003}, but in this application, instead of weighting the neighboring pixels with a multivariate normal distribution, the weights $\gamma_{i,j,c}$ are the posterior probabilities of the BeMM. Therefore, a pixel has a velocity vector for each cloud layer $c$ in a frame. The optimal window size, weighted least-squares regularization, and differential kernel amplitude are: $\mathcal{W} = 16 [ \mathrm{pixels}^2 ]$, $\tau = 1 \times 10^{-8}$, and $\sigma = 1$ respectively. The velocity components in the x-axis are ${\bf U}_c = \{ \text{u}_{i,j,c} \in \mathbb{R} \mid \forall i = 1, \ldots, M, \ \forall j = 1, \ldots, N \}$, and  the velocity components in the y-axis are ${\bf V}_c = \{ \text{v}_{i,j,c} \in \mathbb{R} \mid \forall i = 1, \ldots, M, \ \forall j = 1, \ldots, N \}$. The velocity vectors are in units of pixels per frame, but knowing the geometrical transformation of the frame, they can be transformed to meters per second \cite{TERREN2020a}. The geometric transformation is a function of the Sun's elevation and azimuth angles $\psi : \left( \varepsilon, \alpha \right) \mapsto \Delta \mathbf{x}_{i,j} $ in a frame, it defines the dimensions of a pixel at a given height $ \Delta \mathbf{X} = \{ \left(\Delta x, \Delta y \right)_{i,j} \mid\forall i = 1, \ldots, M, \ \forall j = 1, \ldots, N \}$. This transformation connects the x,y-axis coordinates system with the height, which is the z-axis. The relation holds even when the height is an approximation, since the components of velocity vectors are transformed with respect to the new coordinates system. The velocity vectors of each cloud layer are transformed such as,
\begin{align}
    \label{eq:geometric_transformation}
    u_{i,j} &= \frac{\delta}{f_r}  \cdot \Delta x_{i,j} \sum_{c = 1}^C  \hat{H}_c \cdot \gamma_{i,j,c} \cdot \text{u}_{i,j,c}  \\
    v_{i,j} &= \frac{\delta}{f_r} \cdot \Delta y_{i,j} \sum_{c = 1}^C \hat{H}_c \cdot \gamma_{i,j,c} \cdot \text{v}_{i,j,c}
\end{align}
where $f_r$ is the frame rate of the sequence of images, and $\delta$ is velocity vectors' scale in the WLK approximation.

\subsection{Velocity Vectors Selection}

In order to approximate the potential lines and streamlines of the wind velocity field in a frame, we propose to select the most consistent velocity vectors over a sequence of consecutive frames. The main problems with this approach are that as the vectors are selected over a sequence of images, the amount of vectors is expected to be large; also when optical flow is implemented in dense manner, it yields to noisy vectors. Because of this, we threshold the velocity vectors to reduce both the algorithm's computational burden and the variance of the noise.

\subsubsection{Velocity Vector Segmentation}

The pixel intensity difference between two consecutive frames is computed to find the pixels that show a change. The pixel normalized intensities that were used to compute the velocity vectors are ${\mathbf{I}} = \{ {i}_{i,j} \in \mathbb{R}^{[0, 2^8)} \mid \forall i = 1, \ldots, M, \ \forall j = 1, \ldots, N \}$. The root squared intensity normalized difference is,
\begin{align}
    {d}_{i,j} =\frac{ \sqrt{\left({i}^{k - 1}_{i,j} - {i}^k_{i,j}\right)^2}}{\sum_{i,j}{ \sqrt{\left({i}^{k - 1}_{i,j} - {i}^k_{i,j}\right)^2}}}.
\end{align}

Matrix ${\bf D}$ with normalized differentials $d_{i,j}$ is vectorized and sorted from the lowest to the highest, i.e., ${\mathbf{d}} = \mathrm{sort}(\mathrm{vec} ({\mathbf{D}} ))$. A vector $\bf r$ with the accumulated variance is computed as
\begin{align}
    r_m = \left\{ \sum^m_{i = 1} {\bf d}_i \right\}^{N \cdot M}_{m = 1}.
\end{align}
Then, vector $\mathbf{r}$ is reorganized and set in the original matrix form, defined as $\mathbf{R} = \{ r_{i,j} \in \mathbb{R}^{[0,1)} \mid \forall i = 1, \dots, M, \ \forall j = 1, \dots, N \}$. Finally, a threshold $\tau$ is applied
\begin{align}
    \label{eq:threshold}
    b_{i,j} = 
    \begin{cases}
    1 \quad r_{i,j} \geq \tau \\
    0 \quad \mathrm{Otherwise},
    \end{cases}
\end{align}
where  $\mathbf{B} \in \mathbb{B}$ is a binary image whose elements are 1 when a pixel is selected. The threshold velocity vectors in a frame $k$ are ${{\bf V}'}^{k} = \{ \mathbf{v}^k_{i,j}=\{u_{i,j}^k,v_{i,j}^k\}  \wedge b^k_{i,j} = 1 \mid \forall i = 1, \ldots M, \ \forall j = 1, \ldots, N \}$.

Based on the assumption that a cloud floating in the air follows a trajectory dictated by the wind velocity field, the wind velocity field is approximated using the segmented velocity vectors of $\ell$ last frames. Hence, the set of velocity vectors available to compute the wind velocity field are,
\begin{align}
    \label{eq:dataset}
    \tilde{\mathbf{V}}^{k} = \left[ {\begin{array}{ccc}
       {{\bf V}'}^k \\
       \vdots \\
       {{\bf V}'}^{k - \ell} \\
      \end{array} } \right] \in \mathbb{R}^{2 \times N^k},
\end{align}
the number of samples in $\tilde{\mathbf{V}}^k$ is $N^k$, this number is not the same in each frame $k$. 

\subsubsection{Velocity Vector and Height Distributions}

A velocity vector $\tilde{\mathbf{v}}_i$ (by omitting superindex $k$) in the set $\tilde{\mathbf{V}}^k = \{ \tilde{\mathbf{v}}^k_i \in \mathbb{R}^2 \mid \forall i = 1, \dots, N^k\}$ is assumed to belong to a cloud layer $c$. The probability of a vector to belong to a cloud layer $c$ is modelled as an independent normal random variable $\tilde{\mathbf{v}}_i \sim \mathcal{N}( \boldsymbol{\mu}_c, \boldsymbol{\Sigma}_c )$. The function of the multivariate normal distribution is,
\begin{equation}
    p \left( \tilde{\mathbf{v}}_i \mid  \boldsymbol{\mu}_c, \boldsymbol{\Sigma}_c \right) = \frac{1}{\sqrt{ \left(2 \pi\right)^d \left| \boldsymbol{\Sigma}_c \right| }} \cdot \exp \left\{ - \frac{1}{2} \left( \tilde{\mathbf{v}}_i - \boldsymbol{\mu}_c \right)^\top \boldsymbol{\Sigma}_c^{-1} \left( \tilde{\mathbf{v}}_i- \boldsymbol{\mu}_c \right) \right\}.
\end{equation}
In the case when two cloud layers were detected, we propose to infer the probability distribution of velocity vectors' in each cloud layer with this model,
\begin{align}
    p \left( \tilde{\mathbf{v}}_i \mid \boldsymbol{\Theta} \right) \propto p \left( \tilde{\mathbf{v}}_i \mid \boldsymbol{\mu}_1, \boldsymbol{\Sigma}_1 \right)^{\lambda_i} \cdot  p \left( \tilde{\mathbf{v}}_i  \mid \boldsymbol{\mu}_2, \boldsymbol{\Sigma}_2 \right)^{\left(1 - \lambda_i \right)},
\end{align}
where $\boldsymbol{\Theta} = \{ \boldsymbol{\lambda}, \boldsymbol{\mu}_1, \boldsymbol{\Sigma}_1, \boldsymbol{\mu}_2, \boldsymbol{\Sigma}_2\}$, and $\lambda_i \in \{0, 1\}$.  $\lambda_{i,c}$ is defined as convex, considering that a velocity vector may belong to one or the other wind velocity layer, but no to both. Knowing the vectors that belong to the first cloud layer, the vectors that belong to the second cloud layer are also known, $\lambda_{i,2} = 1 - \lambda_{i,1}$. The lower bound of the data log-likelihood is found applying Jensen's inequality \cite{JENSE1906},
\begin{align}
	\label{eq:velocity_vectors_distribution}
    \log p \left( \tilde{\mathbf{v}}_i \mid {\boldsymbol{\Theta}} \right) \propto \lambda_{i,1} \cdot \log p \left( \tilde{\mathbf{v}}_i \mid {\boldsymbol{\mu}}_1, {\boldsymbol{\Sigma}}_1 \right) + \lambda_{i,2} \cdot \log p \left( \tilde{\mathbf{v}}_i  \mid {\boldsymbol{\mu}}_2, {\boldsymbol{\Sigma}}_2 \right),
\end{align}
so that the posterior distribution is a linear combination of the multivariate normal distributions.

The probabilistic model parameters are inferred using a fixed-point variation of the Iterated Conditional Modes (ICM) \cite{BESAG1986}. The algorithm begins by randomly assigning the velocity vectors to a cloud layer, $\lambda_{i,1} \sim \mathcal{U} ( 0, C - 1 )$. The parameters of velocity vector distributions, in Eq. \eqref{eq:velocity_vectors_distribution} that maximize the data log-likelihood are computed in the first step of the algorithm. These same parameters are used to initialize the inference of the parameters of the height distributions in the second step in Eq. \eqref{eq:height_distributions}. 

In the case of a multivariate normal distribution, the ICM algorithm is iteratively updates parameters. At iteration $t+1$, the means and covariances are,
\begin{align}
    \label{eq:parameters_estimation}
    {\boldsymbol{\mu}}_c^{(t + 1)} = \frac{\sum_i \lambda_{i,c}^{(t)} \cdot \tilde{\mathbf{v}}_i }{\sum_i \lambda_{i,c}^{(t)}}; \ \ \
    {\boldsymbol{\Sigma}}_c^{(t + 1)} = \frac{\sum_{i} \lambda_{i,c}^{(t)} \cdot \left( \tilde{\mathbf{v}}_i  - {\boldsymbol{\mu}}_c^{(t + 1)} \right)^\top \left( {\mathbf{v}}_i  - \tilde{\boldsymbol{\mu}}_c^{(t + 1)} \right)}{\sum_i \lambda_{i,c}^{(t)}}
\end{align}
The vectors are re-assigned to a cloud layer at the end of each parameters update, applying the maximum a posteriori (MAP) criterion
\begin{align}
    \lambda_{i,2}^{(t + 1)} &= \underset{c}{\operatorname{argmax}} \ p \left( \tilde{\mathbf{v}}_i \mid {\boldsymbol{\mu}}_c^{(t + 1)}, {\boldsymbol{\Sigma}}_c^{(t + 1)} \right) - 1 \\
    \lambda_{i,1}^{(t + 1)} &= 1 - \lambda_{i,2}^{(t + 1)}.
\end{align}
After completing the inference of the velocity vectors distribution, it is possible to infer the cloud layer's height using the same method. The velocity vectors in an image were calculated using the WLK method. The algorithm approximates the velocity vector using a set of pixels inside a window. The result is that the velocity vectors do not exactly correspond to a clouds' pixels, which are in motion. Instead, the velocity vectors are assigned to nearby pixels. To identify which layer of clouds $c$, is the highest and which one is the lowest, the height distribution of the pixels is inferred using the MAP classification of the velocity vectors $\mathbf{v}^\prime_{i,j}$ in a image. 

The height of the pixels within the cloud are modelled as independently distributed normal random variables $H_{i,j} \sim \mathcal{N} (\mu_c, \sigma^2_c) $. The probabilistic model to infer the distribution of heights of each cloud layer in a frame is,
\begin{align}
	\label{eq:height_distributions}
    \log p \left( H_{i,j} \mid \boldsymbol{\Theta} \right) \propto \rho_{i,j,1} \cdot \log p \left( H_{i,j} \mid \mu_1, \sigma^2_1 \right) + \rho_{i,j,2} \cdot \log p \left( H_{i,j} \mid \mu_2, \sigma^2_2 \right),
\end{align}
where $\boldsymbol{\Theta} = \{\boldsymbol{P}, \mu_1, \sigma_1, \mu_2, \sigma_2 \}$, and $\rho_{i,j,c} \in \{0, 1\}$ is a convex variable so that $\rho_{i,j,2} = 1 - \rho_{i,j,1}$.

The ICM algorithm is also used to the infer the parameters of the height distributions model. The $\rho_{i,j,c}$ are initialized to the MAP classification of the velocity vectors $\mathbf{v}^\prime_{i,j}$ using the parameters that were inferred using all the velocity vectors $\tilde{\mathbf{v}}_i$ in Eq. \eqref{eq:dataset},
\begin{align}
    \rho_{i,j,2} &= \underset{c}{\operatorname{argmax}} \ p \left( \mathbf{v}^{\prime}_{i,j} \mid \boldsymbol{\mu}_c, \boldsymbol{\Sigma}_c \right) - 1\\
    \rho_{i,j,1} &= 1 - \rho_{i,j,2}.
\end{align}
The parameters of the height distributions are updated using the formulas in Eq. \eqref{eq:parameters_estimation}. The algorithm eventually converges so that the pixels in the frame are segmented where a cloud appears. The segmentation is performed according to the MAP classification of height distributions,
\begin{align}
    \rho_{i,j,2}^{(t + 1)} &= \underset{c}{\operatorname{argmax}} \ p \left( H_{i,j} \mid \mu_c^{(t + 1)}, \sigma_c^{2(t + 1)}\right) - 1\\
    \rho_{i,j,1}^{(t + 1)} &= 1 - \rho_{i,j,2}^{(t + 1)}.
\end{align}
In order to find the height of a given cloud layer, the heights are averaged with this formula,
\begin{align}
    \bar{H}_c = \frac{\sum_{i,j} \rho_{i,j,c} \cdot H_{i,j} \cdot \mathbb{I} \left( b_{i,j} = 1 \right)}{\sum_{i,j} \rho_{i,j,c} \cdot \mathbb{I} \left( b_{i,j} = 1 \right)}.
\end{align}
The wind velocity fields are organized into upper and lower layers by average height $\bar{H}_c$. In this way, each detected wind velocity field has a distribution of velocity vectors, and another distribution of heights.

\subsubsection{Sampling}

In order to reduce the computational burden of the algorithm, a subset of the velocity vectors is selected according to the estimated probability distributions of the vectors. At each layer $c$, we define the importance weights $w_{i,c}^k$ as
 \begin{align}
    w^k_{i,c} \triangleq p \left( \tilde{\mathbf{v}}^k_i \middle| \boldsymbol{\theta}_c \right), \quad w_{i,c}^k \in \mathbb{R}^+.
\end{align}
The weights are normalized to have the characteristics of a probability mass function such as $\sum_{i = 1}^{N_k} \hat w_{i,c}^k = 1$. 

The Cumulative Probability Function (CDF) is computed as
\begin{align}
    \tilde{w}^k_{i,c} = \left\{ \sum_{m = 1}^i \hat{w}^k_{m,c} \right\}_{i = 1}^{N^k}.
\end{align}
In order to select samples for each distribution $p ( \tilde{\mathbf{v}}^k_i \mid \boldsymbol{\theta}_c )$, $N^*/C$ samples are drawn from a uniform distribution,
\begin{align}
    z_{j,c}^k \sim \mathcal{U}\left( 0, 1 \right), \quad j = 1, \ldots, \frac{N^*}{C},
\end{align}
For each value $z_{j,c}^k$, a sample is selected with the criterion 
\begin{align}
    I_{j,c}^k = \operatorname{argmin} \ \mid \tilde{w}_{i,c}^k - z_{j,c}^k \mid, \quad \forall i = 1, \ldots, N_k \quad \forall j = 1, \ldots, \frac{N^*}{C}.
\end{align}
The selected vectors are the ones whose CDF is closest to the values of the uniform samples $z_{j,c}^k$,
\begin{align}
    \tilde{\mathbf{v}}_c^{*k} \triangleq \tilde{\mathbf{v}}^k_{I_{j,c}^k}, \quad \forall j = 1, \ldots, \frac{N^*}{C}.
\end{align}

The subset of selected velocity vectors in frame $k$ for the cloud layer $c$ is $\mathbf{V}_c^{*k} = \{ ( \tilde{u}, \tilde{v} )_{j,c}^{*k} \in \mathbb{R}^2 \mid \forall j = 1, \ldots, N^*/C \} $, the subset of Euclidean coordinate pairs of those selected vectors is $\mathbf{X}_c^{*k} = \{ ( x, y )_{j,c}^{*k} \in \mathbb{N}^2 \mid \forall j = 1, \ldots, N^*/C \}$. 

Assuming that the prior is uniform, the posterior probabilities are,
\begin{align}
    \label{eq:posterior_proba}
    z_{i,c}^{*k} \triangleq \frac{p \left( \tilde{\mathbf{v}}^{*k}_i \middle| \boldsymbol{\theta}_c \right)}{\sum_{c = 1}^C p \left( \tilde{\mathbf{v}}^{*k}_i \middle| \boldsymbol{\theta}_c \right)}.
\end{align}

The sampling is repeated for as many cloud layers detected. All selected subsets of vectors, coordinate pairs, and weights form the dataset that is used to approximate the wind velocity field, 
\begin{align}
    \label{eq:selected_dataset}
    \mathbf{X}^{*k} = 
      \begin{bmatrix}
        x^{*k}_1 & y^{*k}_1 \\
        \vdots & \vdots  \\
        x^{*k}_{N^*} & y^{*k}_{N^*} \\
      \end{bmatrix}, \
    \mathbf{V}^{*k} = 
      \begin{bmatrix}
        \tilde{v}^{*k}_1 & \tilde{u}^{*k}_1 \\
        \vdots & \vdots  \\
        \tilde{v}^{*k}_{N^*} & \tilde{u}^{*k}_{N^*} \\
      \end{bmatrix}, \
    \mathbf{Z}^{*k} = 
      \begin{bmatrix}
        z^{*k}_{1,1} & z^{*k}_{1,c} \\
        \vdots & \vdots  \\
        z^{*k}_{N^*,1} & z^{*k}_{N^*,c} \\
      \end{bmatrix},
\end{align}
where $N^k >> N^*$. 

\section{Flow Visualization}

The atmosphere is a system where the dynamics are continuously changing \cite{KALNAY2003}. This fact implies that a wind velocity field exists, but we can only visualize it where clouds are present. From ground level to the Tropopause, the wind flow can have multiple layers with different velocities in each one of the layers. The wind flow may also be convective, however, for the sake of simplicity, we assume that the multi-layer flow which is observed in IR images is a multi-layer laminar flow. This analysis neither considers the z-component in the motion of a cloud (which is not observable)  nor the possible inter-crossing of cloud layers.

\subsection{Wind Velocity Field Estimation}

Three methods were implemented to estimate the extrapolation function and compare their performances. The first method uses a weighted $\varepsilon$-support vector regression machine  ($\varepsilon$-WSVM) for each one of the velocity components. The second method is a $\varepsilon$-MO-WSVM that estimates both velocity components. The third is an innovation which uses a $\varepsilon$-MO-WSVM with flow constraints ($\varepsilon$-MO-WSVM-FC) to estimate both velocity components. The flow constraints are used to force the extrapolated wind flow to have zero divergence or curl, so it can be assumed that, in the approximated wind flow, streamlines are equivalent to the cloud pathlines.

The regression problem can be formulated as the optimization of a function with the form,
\begin{align}
    f\left(\mathbf{x}_i\right) = \mathbf{w}^\top \varphi \left( \mathbf{x}_i \right) + b,\quad \forall i = 1, \dots, N^*, \quad \mathbf{w}, \mathbf{x}_i \in \mathbb{R}^D,\ b \in \mathbb{R}.
\end{align}
where $\mathbf{x}_i \triangleq \mathbf{x}^{*k}_i$  in our problem, and where $\varphi(\cdot)$ is a transformation into a higher dimensional (possibly infinite) Hilbert space $\mathcal{H}$ endowed with a dot product $\mathcal{K}({\bf x}_i,{\bf x}_j)=\langle \varphi({\bf x}_i),\varphi({\bf x}_j)$. A function  $\mathcal{K}({\bf x}_i,{\bf x}_j)$ is a dot product if it is a bivariate positive semi-definite function that maps ${\bf x}_i,{\bf x}_j$ into $\mathbb{R}$, commonly called a Mercer's kernel or simply a kernel function. 

\subsubsection{Support Vector Machine for Regression}

Assuming $\mathbf{v}_i =\{u_i,v_i\} \triangleq \mathbf{v}_i^{*k}$, the regression problem in a $\varepsilon$-SVM is formulated with an $\varepsilon$-insensitive loss function, which penalizes the errors $ |\varepsilon| > 0$ \cite{DRUCKER1997} for each one of the components in ${\bf v}_i$ and for each cloud layer $c$ as
\begin{align}
    \left| u_i - f \left( \mathbf{x}_i \right)\right|_{\varepsilon} = \max \left[ 0, \left| u_i - f \left( \mathbf{x}_i \right)\right|-\varepsilon \right], \quad \forall i = 1, \dots, N, \quad u_i,\varepsilon \in \mathbb{R},
\end{align}
and identically for $v_i$. The $\varepsilon$-insensitive loss function can be seen as a tube of radius $\varepsilon$ adjusted around the regression hyper-plane via the Support Vectors (SV) \cite{SCHOLKOPF2000}.

The samples are weighted by their probability of belonging to wind velocity field $c$ \cite{ELATTAR2010},
\begin{align}
    z_i &\triangleq z_i^{*k}, \quad z_i \in \mathbb{R}^{\leq 1}. \\
    c_i &= z_i \cdot \frac{C}{N} 
\end{align}
The L2-norm and $\varepsilon$-loss function is applied to the model weights, 
\begin{align}
    \min_{\mathbf{w}, b, \xi, \xi^*} & \quad \frac{1}{2} \| \mathbf{w} \|^2 + \frac{C}{N} \sum_{i = 1}^N z_i \left( \xi_i + \xi_i^* \right) \label{eq:primal}\\ 
    \mathrm{s.t.} & 
    \begin{cases}
    u_{i}-\mathbf{w}^\top \varphi \left( \mathbf{x}_i \right) - b & \leq \varepsilon + \xi_i \\
    \mathbf{w}^\top \varphi \left( \mathbf{x}_i \right) + b  - u_{i} & \leq \varepsilon + \xi_i^* \\
   \xi_i, \xi_i^* & \geq 0
    \end{cases}  \quad i = 1, \ldots, N, \label{eq:primal_constraints}
\end{align}
and identically for $v_i$. The slack variables $\xi_i$ were introduced to relax the constraints of the optimization problem and to deal with unfeasible optimization problems \cite{CORTES1995}. The primal objective function aims to find the trade off between the regularization term $\varepsilon$, the allowed errors or slack variables $\xi_i$ and $\xi_i^*$, and the complexity of the model $c_i$ per weighted sample.

The proposed kernel functions in this analysis are, 
\begin{align}
    \mathcal{K} \left( \mathbf{x}_i, \mathbf{x}_j \right) &= \mathbf{x}_i^\top \mathbf{x}_j, \\ 
    \mathcal{K} \left( \mathbf{x}_i, \mathbf{x}_j \right) &= \exp \left( - \gamma \cdot || \mathbf{x}_i - \mathbf{x}_j ||^2 \right), \\
    \mathcal{K} \left( \mathbf{x}_i, \mathbf{x}_j \right) &= \left( \gamma \cdot \mathbf{x}_i^\top \mathbf{x}_j + \beta \right)^d,
\end{align}
where $\gamma, \ \beta \in \mathbb{R}$, and $d \in \mathbb{N}$ are the kernel hyperparameters that need cross-validation. \cite{SCHOLKOPF2000}. These kernel functions are referred to as linear, radial basis function (RBF) or square exponential, and polynomial of order $d$ respectively \cite{SHAWE2004}.

In order to optimize the constrained problem in functional \eqref{eq:primal} and constraints \eqref{eq:primal_constraints} a Langrangian functional is constructed with the functional and the set of constraints through a dual set of new variables \cite{SMOLA2004}, which leads to a solvable Quadratic Programming problem (QP). The Lagrangian functional is 
\begin{align}
    \mathcal{L} & \left( \mathbf{w}, b,\alpha, \alpha^*, \beta, \xi, \xi^*, \varepsilon, \eta, \eta^* \right) = \\
    & = \frac{1}{2} \| \mathbf{w} \|^2 + c_i \sum_{i = 1}^N \left( \xi_i + \xi_i^* \right) \ \ldots \\
    & - \sum_{i = 1}^N \left( \eta_i \xi_i + \eta_i^* \xi_i* \right) - \sum_{i = 1}^N \alpha_i \left( \varepsilon + \xi_i - u_i + \mathbf{w}^\top \varphi \left( \mathbf{x}_i \right) + b \right) \ \ldots \\
    & - \sum_{i = 1}^N \alpha_i^* \left( \varepsilon + \xi_i^* + u_i -  \mathbf{w}^\top \varphi \left( \mathbf{x}_i \right)- b \right), \quad \forall i = 1, \dots, N, \quad \eta_i, \eta_i^* \in \mathbb{R}.
\end{align}
The derivatives of the primal variables $\mathbf{w},\varepsilon, \xi_i, \xi_i^*$  yield to the following set of equations, which is a case of Karush-Kuhn-Tucker (KKT) conditions,
\begin{align}
    \mathbf{w}^\top &= \sum_{i = 1}^N \left( \alpha_i^* - \alpha_i \right) \varphi \left( \mathbf{x}_i \right), \\
    0 &= \sum_{i = 1}^N \left( \alpha_i - \alpha_i^* \right ), \\
    0 &= c_i - \alpha_{i} - \eta_{i}, \\
    0 &= c_i 
    - \alpha_{i}^* - \eta_{i}^*.
\end{align}
These conditions, together with the complimentary KKT conditions (which force the product of dual parameters $\alpha_i,\alpha^*_i$ with the constraints to be zero) leads to the following dual functional by substitution on the Lagrangian: 
\begin{align}
    \min_{\boldsymbol{\alpha}, \boldsymbol{\alpha}^*} &\quad \frac{1}{2} \cdot \left( \boldsymbol{\alpha} - \boldsymbol{\alpha}^* \right)^\top \mathbf{K} \left( \boldsymbol{\alpha} - \boldsymbol{\alpha}^* \right) + \sum_{i=1}^N \left( {\alpha_i} - \alpha_i^* \right)u_i + \varepsilon \cdot  \mathbf{1}^\top \left( \boldsymbol{\alpha} + \boldsymbol{\alpha}^* \right) \\
    &\mathrm{s.t.}
    \begin{cases} 
        \mathbf{1}^\top \left( \boldsymbol{\alpha} - \boldsymbol{\alpha}^* \right) = 0 \\ 
        0 \leq \alpha_i, \alpha_i^* \leq c_i
    \end{cases} \ \forall i = 1, \dots, N.
\end{align}
where $\mathbf{1}_{1 \times N} = [1, \ldots, 1 ]^\top$, and matrix $\bf{K}$ is a Gram matrix of dot product such that ${\bf K}_{i,j} = \mathcal{K}({\bf x}_i,{\bf x}_j) $. The minimal of the primal function is equivalent to the saddle point on the Lagrangian formulation. The approximated function is,
\begin{align}
    f \left( \mathbf{x} \right) = \sum_{i = 1}^N  \left(\alpha_i  - \alpha^*_{i}\right) \cdot \mathcal{K} \left( \mathbf{x}_i, \mathbf{x}_i \right) + b,
\end{align}
where $b$ is obtained from the complimentary KKT conditions. 

\subsubsection{Multi-Ouput Weighted Support Vector Machine}

When the wind velocity field function is approximated by $\varepsilon$-MO-SVM,the primal regression can be formulated as 
\begin{equation}
    {\bf v}_i = {\bf W}^\top\varphi({\bf x}_i)+{\bf b},
\end{equation}
where each one of the column vectors of primal parameter matrix ${\bf W}$ approximates one of the velocities in vector ${\bf y}_i$. Primal parameters are a function of the dual parameters as well, but the dual parameters $\boldsymbol{\alpha}_i, \boldsymbol{\alpha}_i^{*}$ are vectors in a $2$-dimensional multi-output problem. 

Since independent variables are represented in vectors ${\bf v}_i$, the training set is defined in a vector  $\tilde{\mathbf{v}}_{1 \times 2N}$, and so are the dual parameters $\tilde{\boldsymbol{\alpha}}_{1 \times 2N}$ and $\tilde{\boldsymbol{\alpha}}^*_{1 \times 2N} $ for notation simplicity.

The gram matrix of dot products between input patterns $\varphi({\bf x}_i)$ can be interpreted as the covariance matrix between variables ${\bf v}_i$. Indeed
\begin{equation}
\begin{split}
    \mathbb{E}\left(({\bf v}_i-{\bf b}^\top)({\bf v}_j-{\bf b})\right) &= \mathbb{E}\left(   {\bf W}^\top \varphi({\bf x}_i) \varphi({\bf x}_j)^\top{\bf W}\right) \\
&=    \left(\begin{array}{cc}
    \varphi({\bf x}_i)^\top \Sigma_{11} \varphi({\bf x}_j)&\varphi({\bf x}_i)^\top \Sigma_{12}\varphi({\bf x}_j)\\
      \varphi({\bf x}_i)^\top \Sigma_{21}\varphi({\bf x}_j)&\varphi({\bf x}_i)^\top \Sigma_{22}\varphi({\bf x}_j)\\
    \end{array}\right),
    \end{split}
\end{equation}
where the $2\times2$ covariance $ \mathbb{E}\left({\bf W}{\bf W}^\top\right) $ is interpreted as a model for the dependencies between elements in ${\bf v}_i$, i.e. 
\begin{equation}
    \mathbb{E}\left({\bf W}{\bf W}^\top\right)=
    \left(\begin{array}{cc}
    \Sigma_{11} & \Sigma_{12}\\
    \Sigma_{21} & \Sigma_{22}
    \end{array}\right).
\end{equation}

If we consider that both vertical and horizontal velocities are independent, then $\Sigma_{12}=\Sigma_{21}={\bf 0}$. If we assume further that $\Sigma_{11}=\Sigma_{22}={\bf I}$ for simplicity, which, in turn leads to
\begin{equation}
    \mathbb{E}\left(({\bf v}_i-{\bf b}^\top)({\bf v}_j-{\bf b})\right) =
     \left(\begin{array}{cc}
    \mathcal{K}({\bf x}_i,{\bf x}_j)&0\\
      0&\mathcal{K}({\bf x}_i,{\bf x}_j)\\
    \end{array}\right).
\end{equation}

The Gram matrix $\tilde{\mathbf{K}}_{DN \times DN}$ in the $\varepsilon$-MO-SVM formulation for independent components is,
\begin{align}
    \tilde{\mathbf{K}} = \left( \begin{array}{cc}
        \mathbf{K} & \mathbf{0} \\ 
        \mathbf{0} & \mathbf{K}
    \end{array} \right).
\end{align}
The full kernel matrix in a $\varepsilon$-MO-WSVM is computationally expensive, and it is not implemented in this research.

The extension of weights in the $\varepsilon$-MO-WSVM requires weighting each sample in each output \cite{HAN2014}, 
\begin{align}
    \tilde{z}_i &= \left[ z \ldots z_N \ z_1 \ldots z_N \right]^\top, \quad
    \sum_{i = 1}^{D \cdot N} \tilde{z}_i = 2, \quad \tilde{z}_i \in \mathbb{R}^{\leq 1}.
\end{align}

The dual formulation of the QP problem for the $\varepsilon$-MO-WSVM is,
\begin{align}
    \label{eq:dual_formaulation}
    \min_{\tilde{\boldsymbol{\alpha}}, \tilde{\boldsymbol{\alpha}}^*} & \quad \frac{1}{2} \cdot \left( \tilde{\boldsymbol{\alpha}} - \tilde{\boldsymbol{\alpha}}^* \right)^\top \tilde{\mathbf{K}} \left( \tilde{\boldsymbol{\alpha}} - \tilde{\boldsymbol{\alpha}}^* \right) + \tilde{\mathbf{y}}^\top \left( \tilde{\boldsymbol{\alpha}} - \tilde{\boldsymbol{\alpha}}^* \right) + \varepsilon \cdot \mathbf{1}^\top \left( \tilde{\boldsymbol{\alpha}} + \tilde{\boldsymbol{\alpha}}^* \right) \\
    &\mathrm{s.t.}
    \begin{cases} 
        \mathbf{1}^\top \left( \tilde{\boldsymbol{\alpha}} - \tilde{\boldsymbol{\alpha}}^* \right) = 0 \\ 
        \mathbf{0} \leq \tilde{\alpha}_i, \tilde{\alpha}_i^* \leq \tilde{c}_i \\
    \end{cases}  \ \forall i = 1, \dots, 2 N,
\end{align}
where the extended weighted complexity is $\tilde{c_i} = \tilde{z}_i / 2N$.

\subsubsection{Multi-Ouput Weighted Support Vector Machine with Flow Constraints}

Assuming that the analyzed air parcel is sufficiently small so that the flow can be considered approximately incompressible and irrotational, a new set of flow constraints are added to the original set of constraints with the purpose of visualizing the wind velocity field to force the divergence and the vorticity to zero:
\begin{align}
    \label{eq:flow_constraints}
    \mathrm{s.t.}\begin{cases} 
        \left( \tilde{\mathbf{v}}^{k\top}_c \mathbf{\Delta}_{x,y} \mathbf{\dot{V}} \right)  \cdot \left( \tilde{\mathbf{v}}^{k\top}_c \mathbf{\Delta}_{x,y} \mathbf{\dot{V}} \right)^\top &= 0 \\
        \left(  \tilde{\mathbf{v}}^{k\top}_c \mathbf{\Delta}_{x,y} \mathbf{\dot{D}} \right) \cdot \left( \tilde{\mathbf{v}}^{k\top}_c \mathbf{\Delta}_{x,y} \mathbf{\dot{D}} \right)^\top &= 0,
    \end{cases} 
\end{align}
where the matrices of this expression are defined in Eq. \eqref{eq:matrix_1} Eq. \eqref{eq:set_matrix_1} and Eq. \eqref{eq:set_matrix_2}. To compute the vorticity and divergence, the differentiation of the velocity field along the x-axis, the and y-axis is implemented using operator
\begin{align}
    \label{eq:matrix_1}
    \mathbf{\Delta}_{x,y} =
    \begin{bmatrix}
        \boldsymbol{\Delta}_x & \mathbf{0} \\
        \mathbf{0} & \boldsymbol{\Delta}_y
    \end{bmatrix}_{2N \times 2N},
\end{align}
where the differential operators $\boldsymbol{\Delta}_x$ and $\boldsymbol{\Delta}_y$ are defined as,
\begin{align}
    \label{eq:set_matrix_1}
   \boldsymbol{\Delta}_x =
    \begin{bmatrix}
        -1 & 0 & \ldots & 0\\
        1 & -1 & \ddots & \vdots \\
        0 & 1 & \ddots & 0 \\
        \vdots &  \ddots & \ddots & -1\\
        0 & \ldots & 0 & 1
    \end{bmatrix}_{N \times N}; \quad
    \boldsymbol{\Delta}_y =
    \begin{bmatrix}
        -1 & \ldots & 0 \\
        \vdots & \ddots  & \vdots  \\
        0 & \ldots & -1 \\
        1 & \ldots & 0  \\
        \vdots & \ddots & \vdots \\
        0 & \ldots & 1
    \end{bmatrix}_{N \times N}.
\end{align}
The operators of the velocity field's vorticity and divergence are respectively,
\begin{align}
    \mathbf{\dot{V}} =
    \label{eq:set_matrix_2}
    \begin{bmatrix}
        1 & \ldots & 0 \\
        \vdots & \ddots  & \vdots  \\
        0 & \ldots & 1 \\
        1 & \ldots & 0  \\
        \vdots & \ddots & \vdots \\
        0 & \ldots & 1
    \end{bmatrix}_{2N \times N}; \quad
    \mathbf{\dot{D}} =
    \begin{bmatrix}
        1 & \ldots & 0 \\
        \vdots & \ddots  & \vdots  \\
        0 & \ldots & 1 \\
        -1 & \ldots & 0  \\
        \vdots & \ddots & \vdots \\
        0 & \ldots & -1
    \end{bmatrix}_{2N \times N}.
\end{align}

The velocity field is extrapolated to the entire frame using the inferred parameters in frame $k$
\begin{align}
    \label{eq:extrapolation}
    \hat{\mathbf{v}}^k_c = \left( \tilde{\boldsymbol\alpha}^k_c - \tilde{\boldsymbol\alpha}^{*k}_c \right) \cdot \mathcal{K} \left( \mathbf{X}^{*k}, \mathbf{X} \right) + \mathbf{b}_c^k,
\end{align}
where the velocity components are $\hat{\mathbf{U}}^k_c \triangleq \hat{\mathbf{v}}^k_{x,c}$, and $\hat{\mathbf{V}}^k_c \triangleq \hat{\mathbf{v}}^k_{y,c}$, where  $\hat{\mathbf{U}}_c^k, \hat{\mathbf{V}}^k_c \in \mathbb{R}^{M \times N} $. 

To compute the flow constraints, the velocity field has to be extrapolated to the whole frame using Eq. \eqref{eq:extrapolation}. The constraints in Eq. \eqref{eq:flow_constraints} are added to the constraints in the dual formulation of the $\varepsilon$-MO-SVM in Eq. \eqref{eq:dual_formaulation}.

\section{Wind Velocity Field Dynamics Estimation}

To estimate the wind velocity field dynamics, velocity vectors are approximated using the WLK method. The velocity vectors are weighted by the posterior probabilities of the cloud layers in the image, and transformed to the cloud layer plane. The velocity vectors are segmented and sampled to reduce the noise and the computation burden. The parameters of the $\varepsilon$-MO-WSVM-FC are cross-validated, and the model is trained to estimate the wind velocity field of the detected cloud layers. If an optimal set of parameters is available, it is possible to proceed with the training of $\varepsilon$-MO-WSVM-FM without performing the cross-validation. After training the $\varepsilon$-MO-WSVM-FC, the testing is performed to extrapolate the wind velocity field to the whole image. The streamlines are computed using Eq. \eqref{eq:streamlines} to visualize the trajectory of a cloud. The potential lines are not shown in the Fig. \ref{fig:day_2_layer_0_turbulent}-\ref{fig:day_5_layer_1}, but they are computed with Eq. \eqref{eq:potential_lines}.

The physical process of cloud formation and evolution over time is part of atmospheric thermodynamics and may have divergence and vorticity \cite{LAMB2011}. The air parcel in one frame is very small compared to the whole volume of air contained in the atmosphere. Within this frame we consider it feasible that there is no vorticity or divergence, and the approximated streamlines are equivalent to the pathlines. Henceforth, the obtained results are a numerical approximation of the actual atmospheric air parcel flow.

When we assume that a flow does not have divergence and vorticity, the stream and velocity potential functions are orthogonal, and we can apply Cauchy-Riemann equations to calculate their rates of change \cite{GRANGER1995}. For the stream function we determine $d \phi = v_x dy - v_y dx$ using samples of functions. The trapezoidal rule of numerical analysis is applied to solve the definite integrals \cite{DRAGOMIR1999}. The values of a streamline are,
\begin{align}
    \label{eq:streamlines}
    \boldsymbol{\Phi}_c = \frac{\hat{H}_c}{2} \left[ \left\{ \sum^i_{m = 1} \hat{\bf u}_{m,c} \odot \Delta {\bf y}_{m,c} \right\}^N_{i = 1} - \left\{ \sum^j_{m = 1} \hat{\bf v}_{m,c}  \odot \Delta {\bf x}_{m,c} \right\}^N_{j = 1} \right],
\end{align}
where $\odot$ denotes the element-wise matrix multiplication. This is the sum of element-wise products between each velocity component and its opposite differential increments.

The total change in the potential function is $d\psi = v_x dx + v_y dy$, so we can determine the potential in each pixel $\mathbf{x} = \{x, y\}$ as,
\begin{align}
    \label{eq:potential_lines}
    \boldsymbol{\Psi}_c = \frac{\hat{H}_c}{2} \left[ \left\{ \sum^j_{m = 1} \hat{\bf u}_{m,c} \odot \Delta {\bf x}_{m,c} \right\}^N_{j = 1} + \left\{ \sum^i_{m = 1} \hat{\bf v}_{m,c} \odot \Delta {\bf y}_{m,c}  \right\}^M_{i = 1}  \right].
\end{align}
the sum of each element-wise product between the velocity components, and their differential increment. 

\section{Experiments}

To infer the wind velocity field this method utilizes data acquired by a system that captures circumsolar IR images and measures global solar irradiance using a pyranometer. The IR sensor is a Lepton\footnote{https://www.flir.com/} radiometric camera, which has a wavelength from 8 to 14 $\mu m$ and provides a uniform thermal image. When the radiometric functionality is enabled, the pixels in a frame are turned into temperature measurements in centikelvin units. The resolution of an IR image is $80 \times 60$ pixels, and the diagonal FOV is $60^\circ$. The data is publicly accessible in the Dryad repository \cite{GIRASOL}.

The weather features that were used to compute cloud height as well as to remove cyclostationary artifacts \cite{TERREN2020a} on the IR images are: atmospheric pressure, air temperature, dew point and humidity. The weather station is set to measure every 10 minutes, so the data was interpolated to match the IR images' sampling interval. The weather station is located at the University of New Mexico Hospital, and both its real-time and historical data are publicly accessible\footnote{https://www.wunderground.com/dashboard/pws/KNMALBUQ473}.

\begin{figure}[ht]
    \begin{subfigure}{0.325\linewidth}
        \centering
        \includegraphics[scale = 0.325]{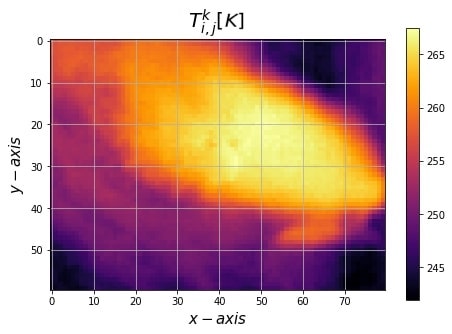}
    \end{subfigure}
    \begin{subfigure}{0.325\linewidth}
        \centering
        \includegraphics[scale = 0.325]{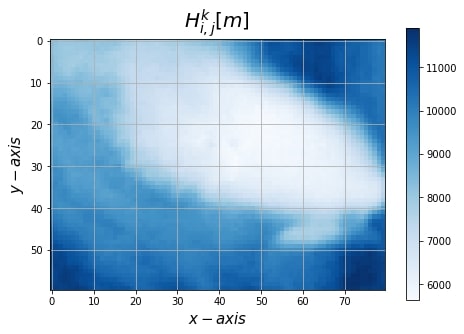}
    \end{subfigure}
    \begin{subfigure}{0.325\linewidth}
        \includegraphics[scale = 0.28]{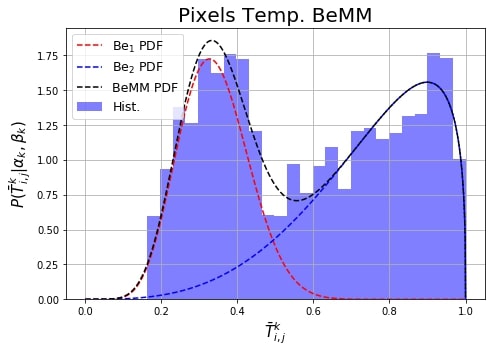}
    \end{subfigure}
    \begin{subfigure}{0.325\linewidth}
        \centering
        \includegraphics[scale = 0.325]{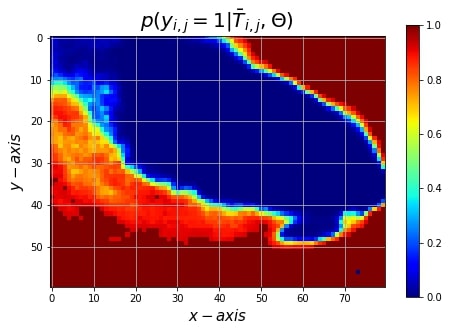}
    \end{subfigure}
    \begin{subfigure}{0.325\linewidth}
        \centering
        \includegraphics[scale = 0.325]{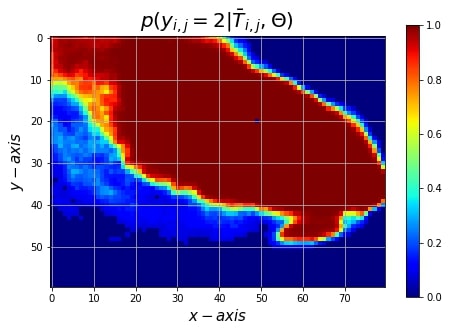}
    \end{subfigure}
    \begin{subfigure}{0.325\linewidth}
        \centering
        \includegraphics[scale = 0.325]{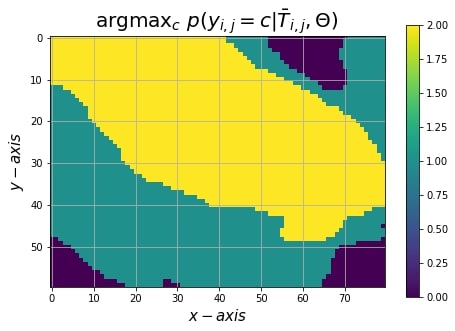}
    \end{subfigure}
\caption{The images in the first row from left to right show the pixel temperatures, pixel heights and the BeMM distribution of the normalized pixel temperatures. The images in the middle row show the posterior probabilities of the upper layer, the posterior probabilities of the lower layer, and MAP classification of the pixels.}
\label{fig:wind_velocity_field}
\end{figure}

The images in the top row of Fig. \ref{fig:wind_velocity_field} show the temperature of the pixels obtained using the radiometric functionality of the IR camera in Kelvin (left pane), the height of the pixels in meters after applying the MALR transformation to the temperatures (center pane), and the temperature $\bar{T}_{i,j}$ histogram in light blue (right pane). Beta distributions in Eq. \eqref{eq:beta_distribution}, are in red and blue, and the BeMM result in Eq. \eqref{EM_likelihood} is in black. The images in the bottom row show the temperature posterior probabilities $\gamma_{i,j,1}$ of the upper layer (left), the temperature posterior probabilities $\gamma_{i,j,2}$ of the lower layer (center), and the MAP classification of the pixels (right). In the image that shows the MAP classification of the pixels, those in dark blue are the segmented pixels that do not belong to a cloud ($b_{i,j} = 0$). 

\subsection{Training Data Construction}

To create a data set for validation purposes we selected 21 consecutive images, per day, on six different days. The images were selected due to the presence of different types of clouds distributed across different heights. The selected images were captured during different seasons and different times of the day. The images from three of the six days show a layer of cirrustratus in winter in the morning, altostratus in spring in the afternoon and stratocumulus in summer in the afternoon. The other three days show two layers of altostratus and stratocumulus in winter at noon, cirrustratus and altocumulus in spring in the afternoon, and  cirruscumulus and cumulus in summer in the morning.
    
The proposed algorithm only requires the validation dataset to be labelled as it is an unsupervised online machine learning algorithm. The validation dataset is used to find the optimal parameters of the algorithm which segments and subsamples the velocity vectors from the last 6 consecutive frames.
    
The targets of the $\varepsilon$-WSVM are the velocity vectors computed using the WLK method. The machine is cross-validated and trained for each frame using the selected velocity vectors of the last 6 frames. The testing error is that of the $\varepsilon$-WSVM approximating the WLK velocity vectors.
    
The average height, velocity magnitude and angle of a cloud was manually calculated for each cloud layer in each sequence of images to use them as ground truth. To do this, the pathline intercepting the Sun was manually segmented. The distance that a cloud has moved in the pathline was calculated in each frame. The height of a cloud layer was measured by segmenting the clouds along the sequence of images. The calculated height, velocity magnitude and angle of each cloud layer was averaged across the validation sequences of images.
    
The wind velocity is a relative measure of the actual velocity in a frame. The algorithm does not need the actual wind velocity. The algorithm requires the height of the clouds to define the space of camera's FOV. The velocity vectors are transformed from pixels per frame to meters per second. Each pixel is projected to its actual dimension in the space defined by the camera's FOV. To know the distance that a cloud will travel in a given time to occlude the Sun, the magnitude of the projected velocity vectors in the space defined by the camera's FOV is calculated. The relative measure of the wind velocity vectors (in pixels per frame) and the height of the clouds (in meters) are connected together in the frames by the geometric transformation. For this reason, the wind velocity that it is required is not the actual wind velocity but the relative wind velocity measured in the frame.

\subsection{Velocity Vectors Calculation, Segmentation and Subsampling Parameters Validation}

The parameters $\delta$ in Eq. \eqref{eq:geometric_transformation}, $\tau$ in Eq. \eqref{eq:threshold}, $\ell$ in Eq. \eqref{eq:dataset} of the velocity vectors selection algorithm were validated using the six sequences of images described above. The velocity estimator was $\varepsilon$-WSVM with a linear kernel. The parameters of $\varepsilon$-WSVM, $\varepsilon$ and $C$, were cross-validated in the same process. The flow velocity constraints were not applied in the validation. 

The average of the approximated wind velocity field height, magnitude and angle was computed, and the Mean Absolute Percentage Error (MAPE) was calculated between the measured and the averaged approximation in each frame. The MAPE was averaged and differentiated across consecutive frames. The combination of parameters that had less averaged MAPE plus total difference of MAPE between consecutive frames was selected. This added difference of MAPE was used to account for the accuracy of the selected model parameters, but also the stability of the models. This stability is optimal if good results are consistently obtained for each one of the images in the same sequence. 

The optimal amplitude $\delta$ of the velocity vector in Eq. \eqref{eq:geometric_transformation}, was found to be $\delta = 2.29$. The optimal threshold $\tau$ in the segmentation of the velocity vector in Eq. \eqref{eq:threshold}, was found to be $\tau = 0.95$.  The optimal number of velocity vectors from $\ell$ last frames to form the dataset in Eq. \eqref{eq:dataset}, was found to be $\ell = 6$. The optimal number of selected samples $N^*$ by sampling algorithm in Eq. \eqref{eq:selected_dataset},  was found that $N^* = 200$ samples are sufficient to perform a robust extrapolation of the wind velocity field to the entire frame.

\begin{figure}[!htb]
    \centering
    \begin{subfigure}{0.475\linewidth}
        \includegraphics[scale = 0.475]{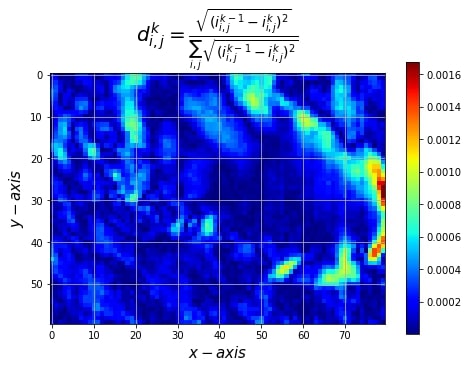}
    \end{subfigure}
    \begin{subfigure}{0.475\linewidth}
        \vspace{16pt}
        \includegraphics[scale = 0.435]{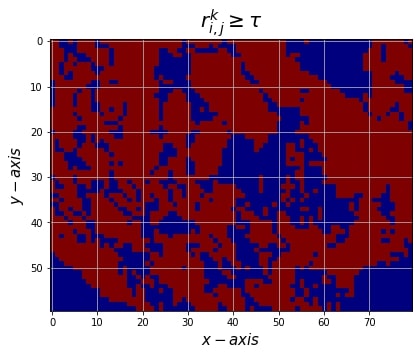}
    \end{subfigure}
\caption{The figures illustrate the proposed method to discriminate between points which probably show a moving cloud air parcel and those that probably show a pixel without movement. The left graph shows the computation of the squared difference between two consecutive frames. The right graph shows in blue those pixels which are considered not moving as their squared difference value is less than a previously validated threshold $\tau$.}
\label{fig:velocity_vectors_seg}
\end{figure}

\begin{figure}[!htb]
    \begin{subfigure}{0.49\linewidth}
        \centering
        \includegraphics[scale = 0.425]{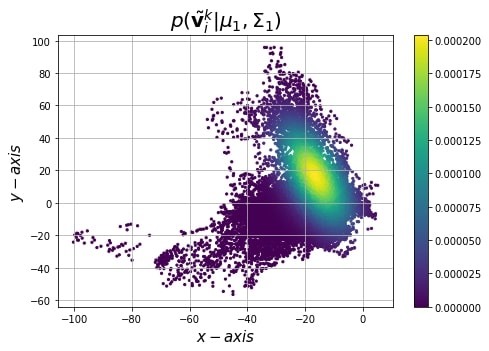}
    \end{subfigure}
    \begin{subfigure}{0.49\linewidth}
        \centering
        \includegraphics[scale = 0.425]{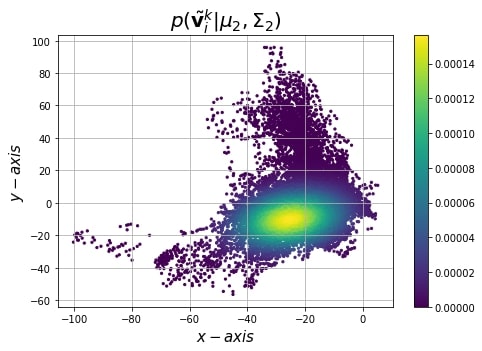}
    \end{subfigure}
    \begin{subfigure}{0.49\linewidth}
        \quad
        \includegraphics[scale = 0.425]{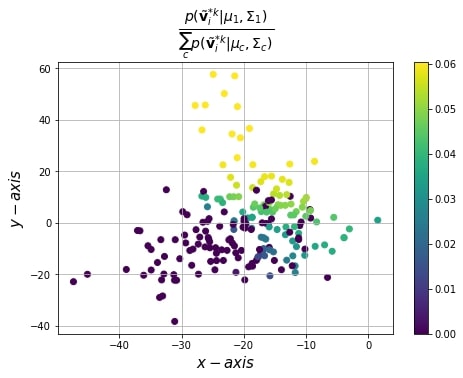}
    \end{subfigure}
    \begin{subfigure}{0.49\linewidth}
        \quad\quad
        \includegraphics[scale = 0.425]{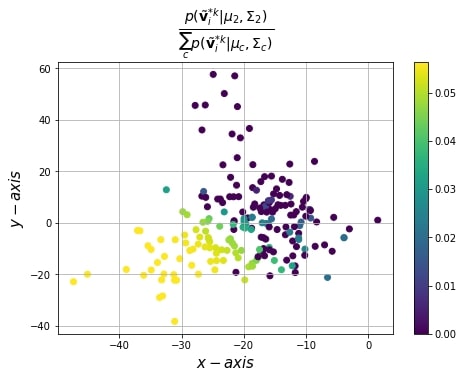}
    \end{subfigure}
\caption{Probability distribution of the velocity vectors and the subsampling implemented to decrease the computational cost. The upper row shows the distribution of the measured velocities represented in $\mathbb{R}^2$. The colormap represents the likelihood of the velocities conditional to a point belonging to the lower layer of clouds (left) and upper layer (right). The lower row shows the downsampled set of vectors (and their posterior probabilities) after applying the downsampling methodology.}
\label{fig:velocity_dist}
\end{figure}

\begin{figure}[!htb]
    \begin{subfigure}{0.475\linewidth}
        \centering
        \includegraphics[scale = 0.425]{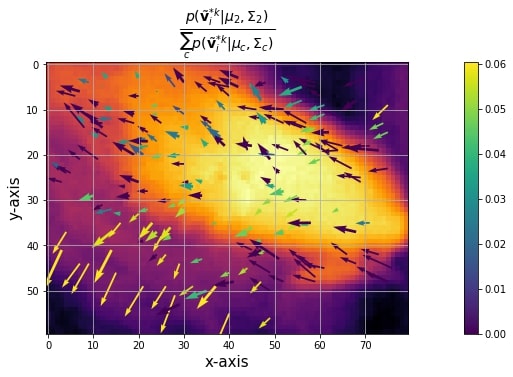}
    \end{subfigure}
    \begin{subfigure}{0.475\linewidth}
        \centering
        \includegraphics[scale = 0.425]{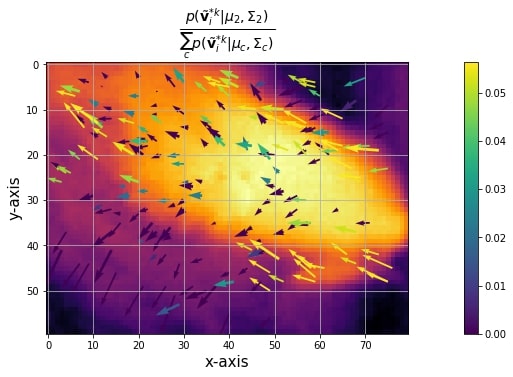}
    \end{subfigure}
\caption{Velocities selected in Fig. \ref{fig:velocity_dist} in their corresponding position. The left image vectors show a color intensity corresponding to their posterior probabilities conditional to the upper cloud layer and the right image to the lower cloud layer. The left image shows in yellow the points clearly belonging to the upper layer, while right image shows in yellow the points that are clearly of the lower layer.} 
\label{fig:velocity_samples}
\end{figure}

The velocity vectors that were segmented in a frame with two layers of clouds are shown in Fig. \ref{fig:velocity_vectors_seg}. The velocity vectors from the last 6 frames after applying the segmentation are shown in the upper row of Fig. \ref{fig:velocity_dist}. In this figure, the colors represent the likelihoods conditional to the upper cloud layer (left), and lower cloud layer (right). The sampled velocity vectors in a frame with two layers of clouds are shown in the bottom row. Fig. \ref{fig:velocity_samples} shows the selected velocities in the bottom row of Fig. \ref{fig:velocity_dist} in their corresponding coordinates. In this figure, the colors represent the posterior probabilities conditional to the upper cloud layer (left), and lower cloud layer (right).
 
\subsection{$\varepsilon$-MO-WSVM-FC Parameters Validation}

After optimal values of $\delta$, $\tau$ and $\ell$ have been chosen, the parameters of the proposed $\varepsilon$-MO-WSVM-FC are cross-validated using the validation data or an online ML approach. This means that the experiments with the $\varepsilon$-MO-WSVM-FC have two different setups. In the first, the parameters are cross-validated in each testing frame. In the second, the parameters are fixed to the optimal values obtained in a previous cross-validation using the validation data. This is done to check for the validity of the previously obtained parameters, which would significantly reduce the velocity field estimation computational burden.

The objective of the constraints is that the divergence and vorticity of approximated velocity field are zero in the clouds' plane. The velocity fields shown in Fig. \ref{fig:day_2_layer_0} to \ref{fig:day_5_layer_1} have divergence and vorticity after the field is projected to the camera plane. This is caused by the non-linear geometric transformation applied to the velocity vectors.

\subsection{Wind Velocity Field Estimation with New Data}

Unlike the $\varepsilon$-MO-WSVM-FC, the experiments with the $\varepsilon$-WSVM and $\varepsilon$-MO-WSVM use only the first setup. These models are validated and trained for each testing frame. The results are compared with a Gaussian process for regression (GPR) for each one of the velocity components \cite{RASMUSSEN2005}, a Multi-Output Ridge Regression (MO-RR) with independent components (which is a special case of Tikhonov's regularization \cite{TIKHONOV1977}) and a multi-output Gaussian process for regression (MO-GPR) with correlation between velocity components \cite{BONILLA2008}.

The testing data is composed of sequences of 28 images from 10 different days. The sequences are from different seasons and different times of the day. Five days have one velocity field layer and the other five have two layers. 75\% of this data is chosen for the online training and validation of the models. The rest of data is used for testing. The testing set is from $k + \ell$ frames ahead of the training set from frame $k$. The number of frames ahead is equal to the lag of the velocity vectors in the data $\ell = 6$. The methodologies implemented in the validation are the standard grid search and 3-fold cross-validation. The parameters cross-validated in the $\varepsilon$-WSVM, $\varepsilon$-MO-WSVM and $\varepsilon$-MO-WSVM-FC are $C$ and $\varepsilon$. The MO-RR requires the cross-validation of the regularization parameter. In the GPR and MO-GPR the parameters are found by numerical gradient, optimizing the marginal log-likelihood. The kernel functions are: linear, RBF, polynomial of order two ($\mathcal{P}^2$), and polynomial of order three ($\mathcal{P}^3$). The optimal parameters for the $\varepsilon$-MO-WSVM-FC are displayed in Table \ref{tab:mo_ewsvm_fc}.

\begin{figure}[htb!]
    \begin{subfigure}{\linewidth}
    \centering
    \includegraphics[scale = 0.21]{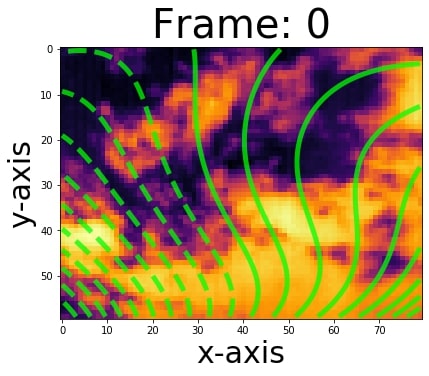}
    \includegraphics[scale = 0.21]{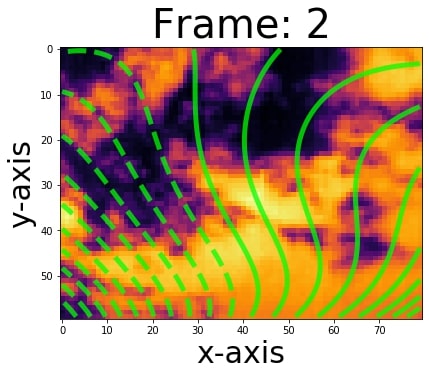}
    \includegraphics[scale = 0.21]{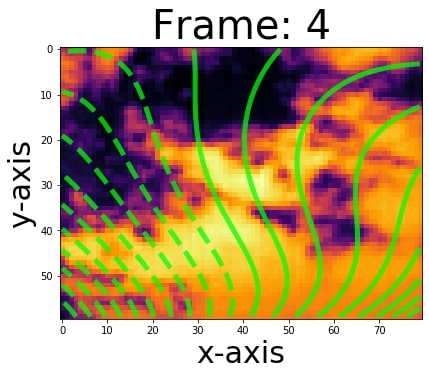}
    \includegraphics[scale = 0.21]{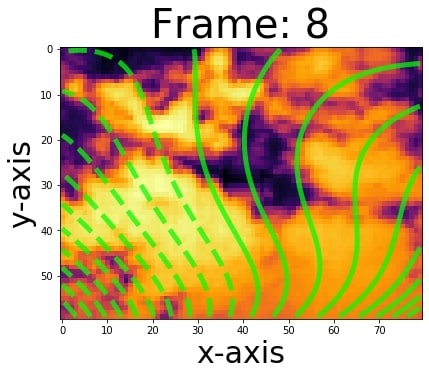}
    \includegraphics[scale = 0.21]{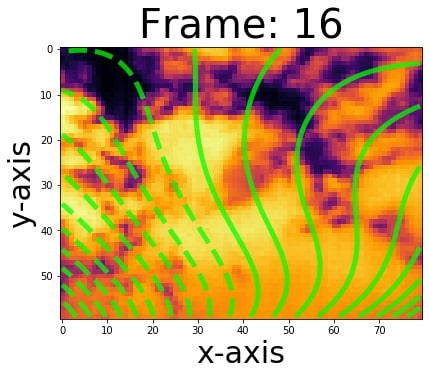}
    \caption{Streamlines approximated using the $\varepsilon$-MO-WSVM with a $\mathcal{P}^3$ kernel. }
    \label{fig:day_2_layer_0_turbulent}
    \end{subfigure}
    \begin{subfigure}{\linewidth}
    \centering
    \includegraphics[scale = 0.21]{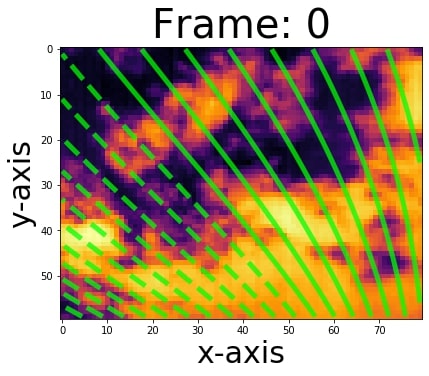}
    \includegraphics[scale = 0.21]{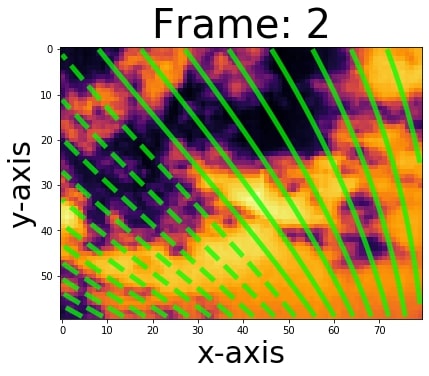}
    \includegraphics[scale = 0.21]{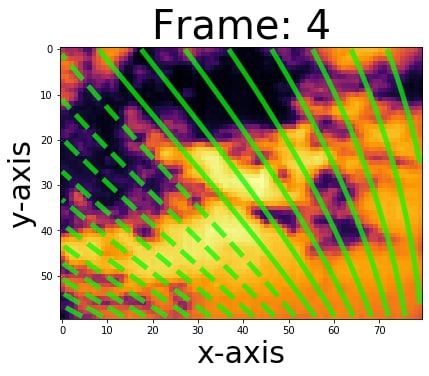}
    \includegraphics[scale = 0.21]{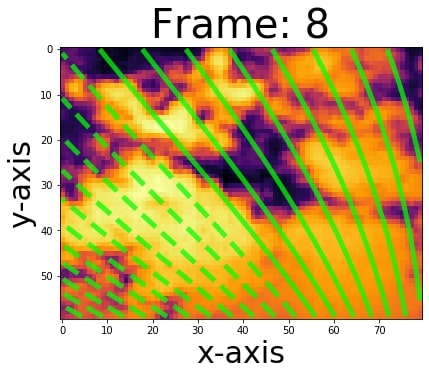}
    \includegraphics[scale = 0.21]{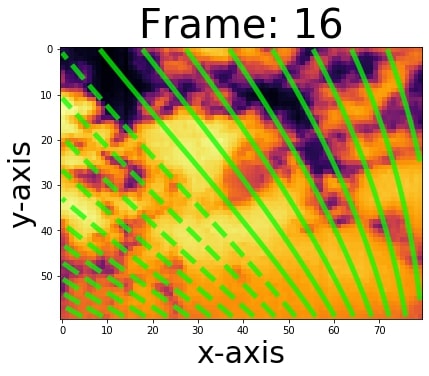}
    \caption{Streamlines approximated using the $\varepsilon$-MO-WSVM-FC with a linear kernel.}
    \label{fig:day_2_layer_0}
    \end{subfigure}
\caption{Comparison between (a) the streamlines computed by a standard Multioutput WSVM  ($\varepsilon$-MO-WSVM) and (b) the introduced WSVM with divergence and vorticity constraints ($\varepsilon$-MO-WSVM-FC) in a sequence of images with  elevation: $46.74^\circ$ and azimuth: $174.21^\circ$. The  images are organized chronologically from the left to the right. The time between frames is 15 s. The distance in the sequences across time is: 0 s, 30 s, 1 min 2 min 4 min. The sequence shows a day when a single cloud layer was detected.  The top sequence visualizes a non-realistic approximation of the flow. A compression is induced to the gas in the bottom left of the frame, and an expansion is induced in the top right of the frame.}
\end{figure}

\begin{figure}[htb!]
    \centering
    \begin{subfigure}{\linewidth}
    \centering
    \includegraphics[scale = 0.21]{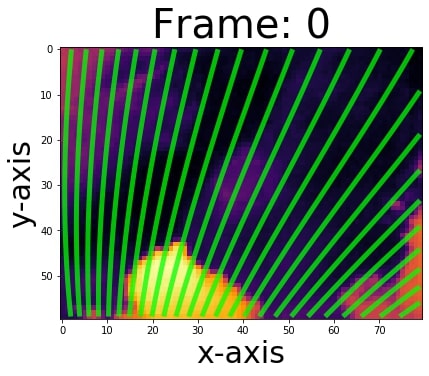}
    \includegraphics[scale = 0.21]{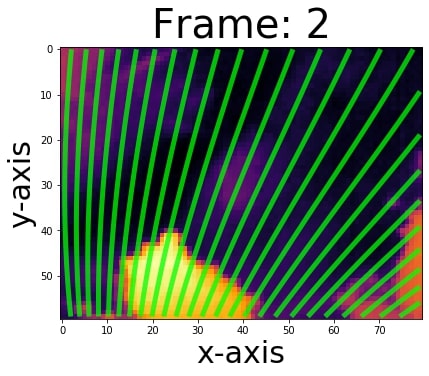}
    \includegraphics[scale = 0.21]{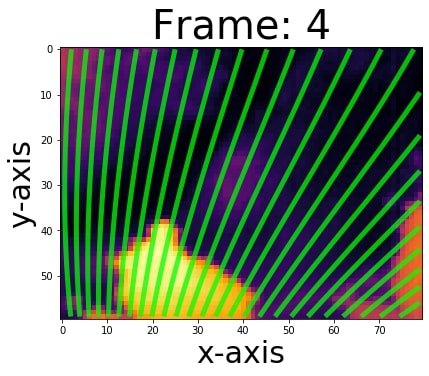}
    \includegraphics[scale = 0.21]{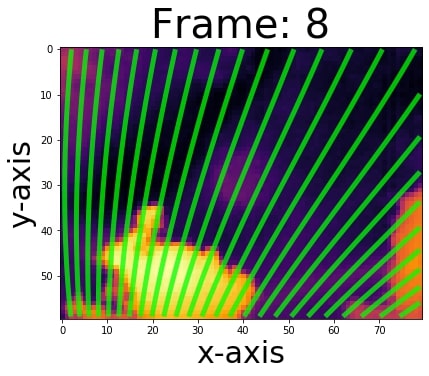}
    \includegraphics[scale = 0.21]{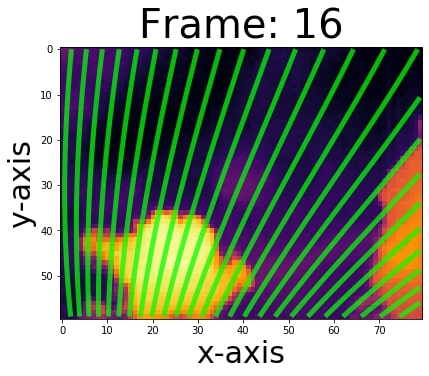}
    \caption{Flow visualization of the upper cloud layer in day 2. Elevation: $55.66^\circ$; azimuth: $200.35^\circ$.}
    \label{fig:day_3_layer_0}
    \end{subfigure}
    \begin{subfigure}{\linewidth}
    \centering
    \includegraphics[scale = 0.21]{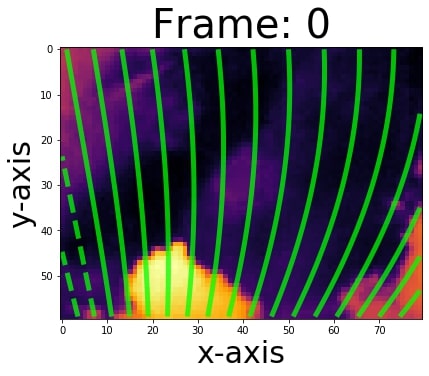}
    \includegraphics[scale = 0.21]{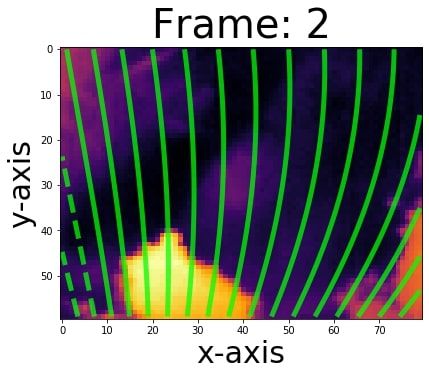}
    \includegraphics[scale = 0.21]{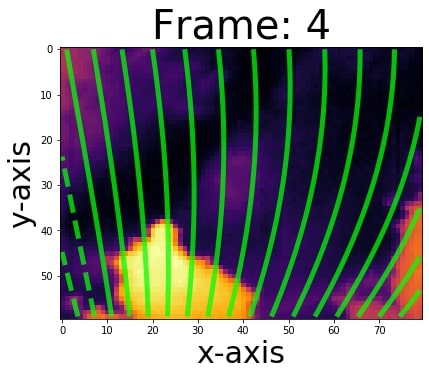}
    \includegraphics[scale = 0.21]{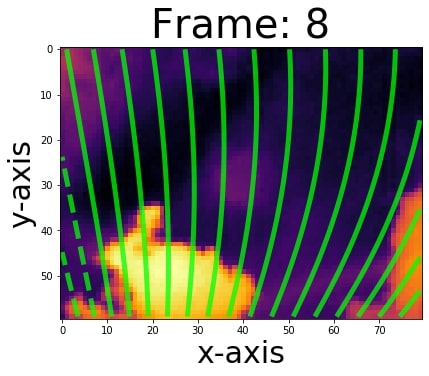}
    \includegraphics[scale = 0.21]{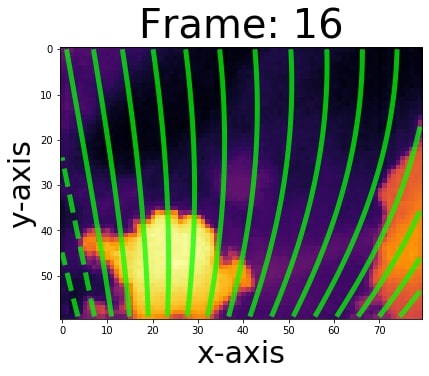}
    \caption{Flow visualization of the lower cloud layer in day 2.}
    \label{fig:day_3_layer_1}
    \end{subfigure}
    \begin{subfigure}{\linewidth}
    \centering
    \includegraphics[scale = 0.21]{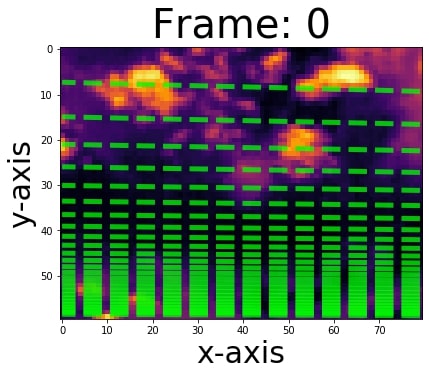}
    \includegraphics[scale = 0.21]{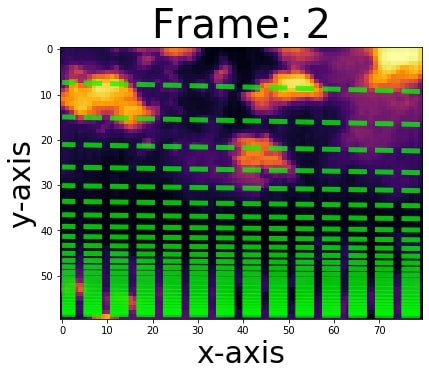}
    \includegraphics[scale = 0.21]{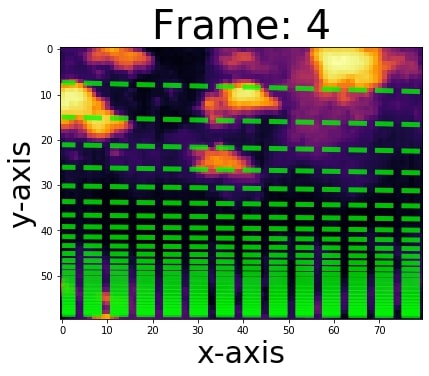}
    \includegraphics[scale = 0.21]{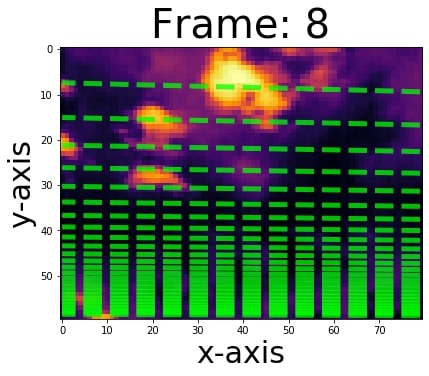}
    \includegraphics[scale = 0.2]{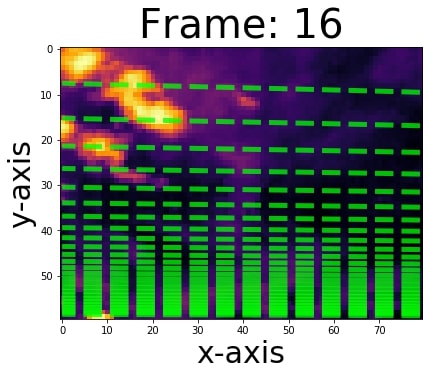}
    \caption{Flow visualization of the upper cloud layer in day 4. }
    \label{fig:day_4_layer_0}
    \end{subfigure}
    \begin{subfigure}{\linewidth}
    \centering
    \includegraphics[scale = 0.21]{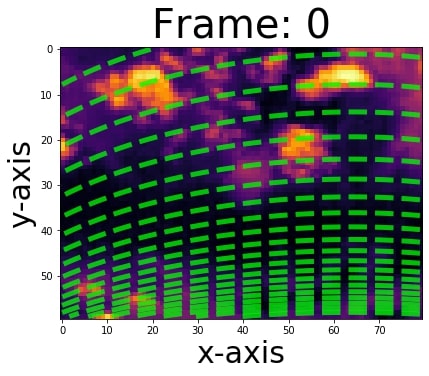}
    \includegraphics[scale = 0.21]{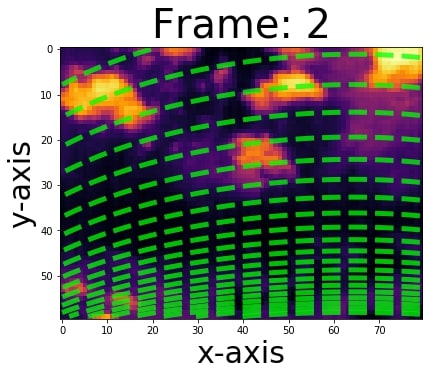}
    \includegraphics[scale = 0.21]{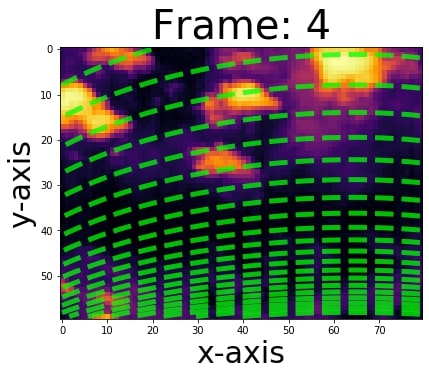}
    \includegraphics[scale = 0.21]{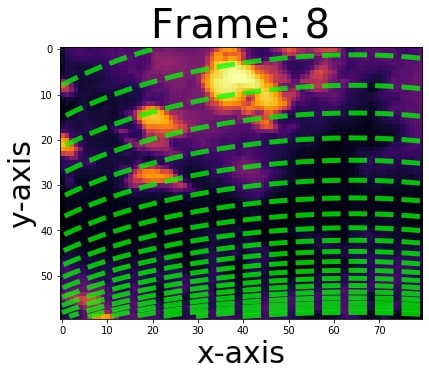}
    \includegraphics[scale = 0.21]{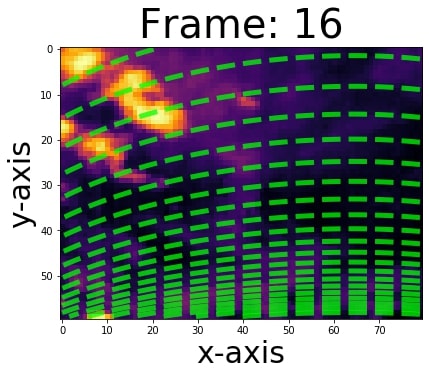}
    \caption{Flow visualization of the lower cloud layer in day 3.}
    \label{fig:day_4_layer_1}
    \end{subfigure}
    \begin{subfigure}{\linewidth}
    \centering
    \includegraphics[scale = 0.21]{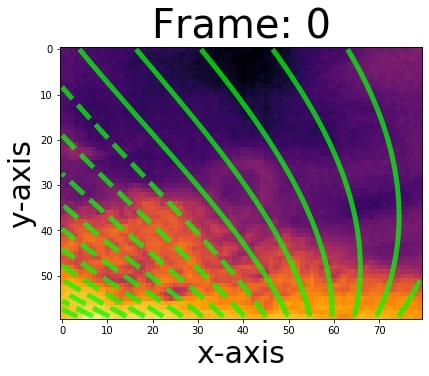}
    \includegraphics[scale = 0.21]{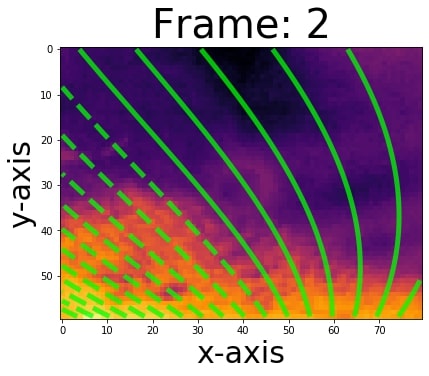}
    \includegraphics[scale = 0.21]{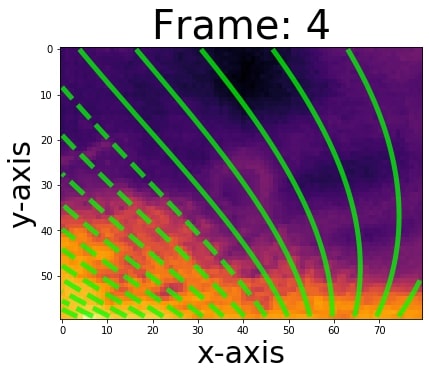}
    \includegraphics[scale = 0.21]{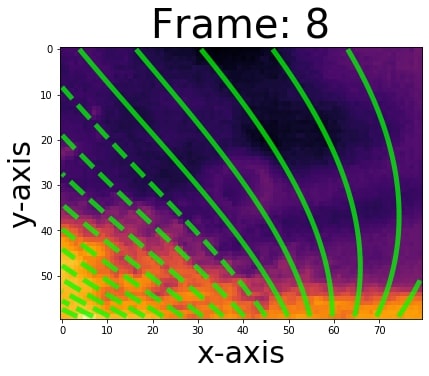}
    \includegraphics[scale = 0.21]{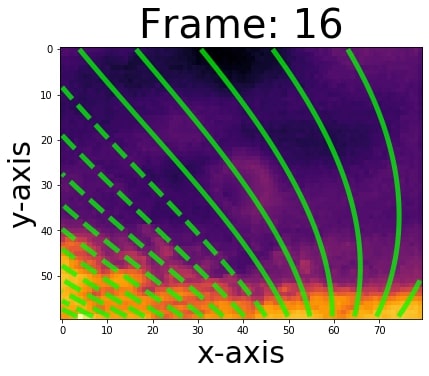}
    \caption{Flow visualization of the upper cloud layer in day 4. }
    \label{fig:day_5_layer_0}
    \end{subfigure}
    \begin{subfigure}{\linewidth}
    \centering
    \includegraphics[scale = 0.21]{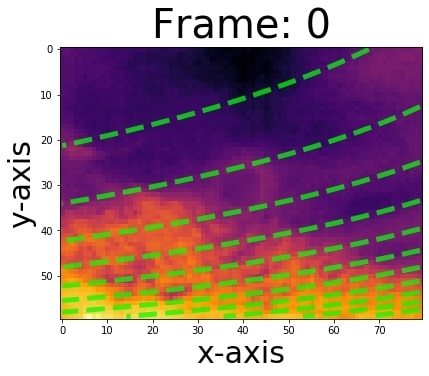}
    \includegraphics[scale = 0.21]{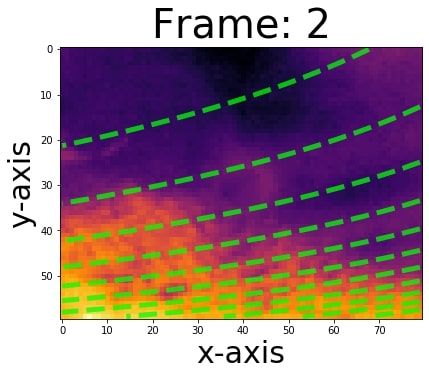}
    \includegraphics[scale = 0.21]{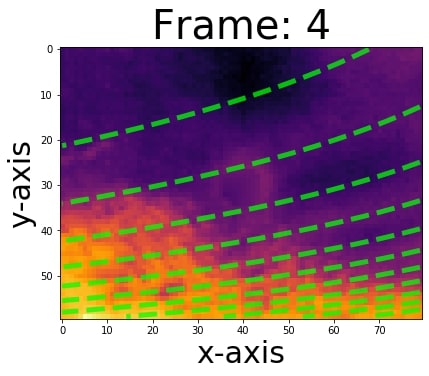}
    \includegraphics[scale = 0.21]{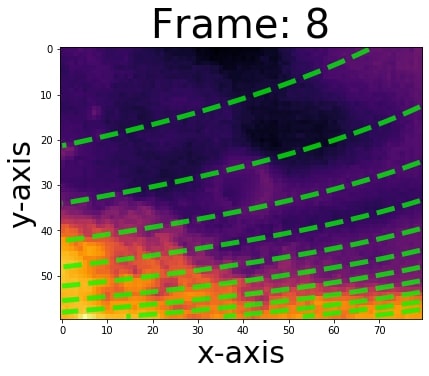}
    \includegraphics[scale = 0.21]{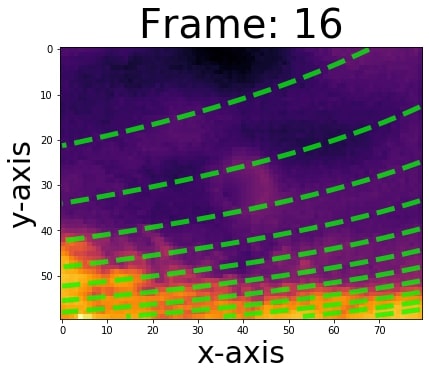}
    \caption{Flow visualization of the lower cloud layer in day 4.}
    \label{fig:day_5_layer_1}
    \end{subfigure}
\label{fig:multilayer_flow}
\caption{Streamlines approximated using the $\varepsilon$-MO-WSVM-FC with a linear kernel. Elevation: $32.15^\circ$; azimuth: $180.29^\circ$. The sequence of IR images are organized chronologically from the left to the right similar to in Fig. \ref{fig:day_2_layer_0_turbulent} and \ref{fig:day_2_layer_0}. The displayed sequences  are from days when two different cloud layers were detected. The upper layer of clouds is displayed in the top row of the sequence, the bottom row displays the lower layer.}
\end{figure}

The criteria for selecting the most suitable model and kernel function for our application is a trade-off between divergence and vorticity, Weighted Mean Absolute Error (WMAE), and the computing time. The values of these metrics are summarized for the models without constraints in Table \ref{tab:wsvm_fc}, and for the $\varepsilon$-MO-WSVM-FC in Table \ref{tab:mo_ewsvm_fc} . The experiment of the $\varepsilon$-MO-WSVM without flow constraints using a $\mathcal{P}^3$ kernel is shown in Fig. \ref{fig:day_2_layer_0_turbulent}, and that same experiment implemented with the $\varepsilon$-MO-WSVM-FC using a linear kernel is in Fig. \ref{fig:day_2_layer_0}. In sequences of images in which two layers of clouds were detected, the experiments of the $\varepsilon$-MO-WSVM-FC using a linear kernel to approximate wind velocity field are shown in the Fig. \ref{fig:day_3_layer_0} to \ref{fig:day_5_layer_1}.

The experiments were carried out in the Wheeler high performance computer of UNM-CARC, which uses SGI AltixXE Xeon X5550 at 2.67GHz with 6 GB of RAM memory per core, 8 cores per node, 304 nodes total, and runs at 25 theoretical peak FLOPS. It has Linux CentOS 7 installed. The DAQ is localized on the roof of UNM-ME building in Albuquerque, NM.
 
\begin{table}[!htb]
\centering
\small
\setlength{\tabcolsep}{6.pt} 
\renewcommand{\arraystretch}{1.25} 
\begin{tabular}{lccccc}
\toprule
\multicolumn{6}{c}{$\varepsilon$-WSVM} \\
\midrule
$\mathcal{K} \left(\mathbf{x}, \mathbf{x}^*\right) $ & MAE & WMAE & $\nabla \cdot \vec{V}$ & $\nabla \times \vec{V}$ & Time [s] \\
\midrule
Linear & 13.37 & 12.55 & 1.69$\cdot 10^3$ & 2.17$\cdot 10^3$ & 90.01 \\
RBF & 13.39 & 12.61 & 6.25$\cdot 10^3$ & 6.40$\cdot 10^3$ & 365.72 \\
$\mathcal{P}^2$ & 14.06 & 13.22 & 1.20$\cdot 10^4$ & 1.19$\cdot 10^4$ & 2413.79 \\
$\mathcal{P}^3$ & 14.90 & 13.95 & 8.98$\cdot 10^4$ & 9.48$\cdot 10^4$ & 3468.75 \\
\midrule
\multicolumn{6}{c}{$\varepsilon$-MO-WSVM} \\
\midrule
Linear & 13.27 & \textbf{12.49} & 1.30$\cdot 10^3$ & 1.35$\cdot 10^3$ & 162.70 \\
RBF & 14.00 & 13.13 & \textbf{1.21}$\mathbf{\cdot 10^3}$ & \textbf{1.22}$\mathbf{\cdot 10^3}$ & 560.54 \\
$\mathcal{P}^2$ & 14.25 & 13.53 & 1.43$\cdot 10^4$ & 1.71$\cdot 10^4$ & 5635.31 \\
$\mathcal{P}^3$ & 19.29 & 18.12 & 8.89$\cdot 10^5$ & 8.92$\cdot 10^5$ & 7284.54 \\
\midrule
\multicolumn{6}{c}{GPR} \\
\midrule
Linear & 12.56 & 12.56 & 2.62$\cdot 10^3$ & 3.27$\cdot 10^3$ & 6.50 \\
RBF & 12.89 & 12.88 & 1.24$\cdot 10^4$ & 1.27$\cdot 10^4$ & 6.43 \\
$\mathcal{P}^2$ & 12.52 & 12.50 & 7.27$\cdot 10^3$ & 9.02$\cdot 10^3$ & 6.44 \\
$\mathcal{P}^3$ & 12.67 & 12.68 & 2.72$\cdot 10^4$ & 3.11$\cdot 10^4$ & \textbf{6.42} \\
\midrule
\multicolumn{6}{c}{MO-RR} \\
\midrule
Linear & 12.62 & 12.58 & 2.62$\cdot 10^3$ & 3.31$\cdot 10^3$  & 6.71 \\
RBF & 13.43 & 13.35 & 3.95$\cdot 10^3$ & 7.24$\cdot 10^3$ & 11.80 \\
$\mathcal{P}^2$ & 12.55 & 12.55 & 1.53$\cdot 10^4$ & 1.15$\cdot 10^4$ & 29.76 \\
$\mathcal{P}^3$ & 12.70 & 12.64 & 3.17$\cdot 10^5$ & 2.20$\cdot 10^5$ & 41.16 \\
\midrule
\multicolumn{6}{c}{MO-GPR} \\
\midrule
Linear & 12.57 & 12.58 & 2.69$\cdot 10^3$ & 3.34$\cdot 10^3$ & 8.07 \\
RBF & 12.81 & 12.80 & 1.21$\cdot 10^4$ & 1.23$\cdot 10^4$ & 17.67 \\
$\mathcal{P}^2$ & 12.53 & 12.55 & 1.10$\cdot 10^4$ & 1.09$\cdot 10^4$ & 11.31 \\
$\mathcal{P}^3$ & 12.54 & 12.55 & 4.12$\cdot 10^4$ & 4.49$\cdot 10^4$ & 11.19 \\
\bottomrule
\end{tabular}
\caption{The table above shows the testing results of the different kernel learning methods without flow constraints. The fist method is the $\varepsilon$-WSVM-FC, the second is $\varepsilon$-MO-WSVM, the third is the GPR, the fourth is the MO-RR and the fifth is the MO-GPR. The wind velocity fields approximated by all the methods have low MAE and WMAE with high divergence and vorticity. The fastest methods are GPR and MO-GPR as the optimization of the parameters is performed via numerical gradient.}
\label{tab:wsvm_fc}
\end{table}

\begin{table}[!htb]
\centering
\small
\setlength{\tabcolsep}{3.75pt}
\renewcommand{\arraystretch}{1.25}
\begin{tabular}{lccccccccccccc}
\toprule
\multicolumn{14}{c}{$\varepsilon$-MO-WSVM-FC} \\
\midrule
{} & \multicolumn{4}{c}{Optimal Parameters} & \multicolumn{4}{c}{Online Parameters Cross-Validation} & \multicolumn{5}{c}{Fixed Optimal Parameters} \\
$\mathcal{K} \left(\mathbf{x}, \mathbf{x}^*\right) $ & $C$ & $\varepsilon$ & $\gamma$ & $\beta$ & MAE & WMAE & $\nabla \cdot \vec{V}$ & $\nabla \times \vec{V}$ & MAE & WMAE & $\nabla \cdot \vec{V}$ & $\nabla \times \vec{V}$ & Time [s] \\
\midrule
Linear & 38.50  & 0.19 & & & 14.22 & \textbf{13.36} & \textbf{0.0} & \textbf{0.0} & 14.24 & 13.35 & \textbf{0.0} & \textbf{0.0} & 58.54 \\
RBF & 38.52 & 0.35 & 13.92 & & 14.55 & 13.53 & 30.96 & 30.86 & 14.12 & \textbf{13.05} & 136.97 & 138.80 & 114.71 \\
$\mathcal{P}^2$ & 39.72 & 0.24 & 3.78 & 44.8 & 14.36 & 13.48 & 77.22 & 70.85 & 14.48 & 13.59 & 30.44 & 30.77 & 130.92 \\
$\mathcal{P}^3$ & 12.88 & 0.22 & 5.61 & 8.34 & 15.34 & 14.34 & 1.74$\cdot 10^4$ & 1.66$\cdot 10^4$ & 45.03 & 44.48 & 2.19$\cdot 10^6$ & 1.97$\cdot 10^6$ & 145.50 \\
\bottomrule
\end{tabular}
\caption{This table shows the optimal sets of parameters obtained cross-validating and training the $\varepsilon$-MO-WSVM-FC in each training image, the results cross-validating the parameters and training the $\varepsilon$-MO-WSVM-FC in each testing image, and the testing results training the $\varepsilon$-MO-WSVM-FC in each testing image using the optimal sets of parameters previously cross-validated in the training data.}
\label{tab:mo_ewsvm_fc}
\end{table}

\section{Discussion}

This investigation adds new insights into the computational methods to forecast the trajectory of clouds and predict the occlusion of the Sun. The proposed method visualizes the wind velocity field using IR images of clouds. The algorithm introduced here differs from previous investigations in that it is based on fluid dynamics. The experiments show that the pathlines are equivalent to the streamlines in images where is possible to extract  enough information about the wind flow from the clouds.
 
From the summary of the experiments presented in Table \ref{tab:wsvm_fc} and \ref{tab:mo_ewsvm_fc}, several aspects can be highlighted. The overall performance of the $\varepsilon$-SVM increases when the samples are weighted, since the weights represent the probability of the vector to belong to the corresponding layer. Vectors with a very low probability do not contribute to the solution. Furthermore, the computing time of the $\varepsilon$-WSVM is lower than the $\varepsilon$-MO-WSVM as the Gram matrix dimensions are smaller. The flow divergence and vorticity are negligible when they are approximated using the $\varepsilon$-MO-WSVM-FC, but the computing time is larger. The results are similar between the three models but the $\varepsilon$-MO-WSVM and $\varepsilon$-MO-WSVM-FC models tend to show better performance.

The best result without cross-validation in WMAE is obtained by the $\varepsilon$-MO-WSVM-FC with RBF kernel (see Table \ref{tab:mo_ewsvm_fc}). The flow approximated by this model has very low vorticity and divergence, which means that the pathlines and streamlines are approximately equivalent. When a trade-off is considered between vorticity, divergence, WMAE, and computing time, the most promising models are the $\varepsilon$-MO-WSVM-FC with linear kernel and RBF kernel.  The computing time required for the linear kernel is lower, as the kernel does not have hyperparameters, but it is still high for a real-time application. Vorticity and divergence are removed in the approximated flow. On the other hand, the $\varepsilon$-WSVM with linear kernel, which has not flow constraints, is feasible in real-time application but the approximated flow is turbulent. When the pathlines begin to be the same as  the streamlines, the flow constraints can be relaxed to reduce the vorticity and divergence within a feasible computing time.

In the implementation of the algorithm, the process of cross-validating the parameters of the $\varepsilon$-MO-WSVM-FC is computationally expensive, and the kernels may have hyperparameters which also require cross-validation. However, the optimal set of parameters is nearly identical during sort sequences. We propose to implement an exhaustive cross-validation in parallel with running the algorithm. This provides a pre-computed set of parameters for the $\varepsilon$-MO-WSVM-FC and the kernel function that can be used in the consecutive images until the online cross-validation is finished.

\section{Conclusions}

This article introduces a method to visualize wind velocity fields using physical features extracted from infrared images of clouds. The images are recorded using a ground-based infrared camera mounted on a solar tracker that maintains the Sun in the center of the images. The velocity vectors are transformed from the Euclidean frame of reference to the infrared camera non-linear frame of reference. The wind velocity field estimation is based on unsupervised online machine learning methods that independently infer the distribution of the velocity vectors and the height of the clouds. Segmenting and subsampling the velocity vectors provides a computationally tractable solution. The wind velocity field is extrapolated to the entire frame using only information extracted from a cloud. This is achieved with the use of a $\varepsilon$-MO-WSVM which includes flow constraints in the quadratic programming problem formulation.

The methods to compute the motion vectors produce a noisy approximation of the velocity vectors in the frame. It is possible to improve the quality of the velocity vectors adding weights to the least-squares solution in the Lucas-Kanade, and later segmenting the velocity vectors. Once the noise is reduced, a subsample of vectors is sufficient to approximate the velocity field in the entire frame. This makes a real-time implementation of the algorithm for wind flow visualization feasible. This is important, because the wind velocity field visualization predicts the pathlines of the clouds. The extrapolation of the wind velocity field to the entire frame is useful to anticipate where a cloud will be, or where it may appear in the frame. Additional constraints in the SVM yields better results, approximating the wind velocity field in the infrared images.

Further research in this area may focus on predicting the occlusion of the Sun or the attenuation of solar irradiance using the streamline (i.e. pathline) that intercepts the Sun, and the magnitude of the wind velocity field in this streamline. The prediction of the wind velocity field distribution across space and time using Bayesian regression methods is suitable for the selection of the most likely intercepting streamlines. Forecasting solar irradiance is out of the scope of this paper. The prediction methods could use the accumulated distance of the pixels along the streamline (starting from the Sun) divided by the averaged magnitude of the approximated velocity vectors, to estimate the arrival time of the air parcel in that pixel. In the existing literature there are multiple optimization algorithms that might speed up the convergence of the $\varepsilon$-MO-WSVM-FC. These methods may increase the accuracy of very short-term solar irradiance forecasting algorithms that are necessary to optimize the dispatch and storage of energy in power grids that used solar resources.

\section{Acknowledgments}

This work has been supported by NSF EPSCoR grant number OIA-1757207 and the King Felipe VI endowed Chair. Authors would like to thank the UNM Center for Advanced Research Computing, supported in part by the National Science Foundation, for providing the high performance computing and large-scale storage resources used in this work.

\bibliographystyle{unsrt}  
\bibliography{mybibfile}

\end{document}